\documentclass[american]{elsarticle}
\usepackage[T1]{fontenc}
\usepackage{textcomp}
\usepackage[utf8]{inputenc}
\usepackage{bm}
\usepackage{amsmath}
\usepackage{amssymb}
\usepackage{graphicx}
\usepackage{esint}

\makeatletter
\journal{Annals of Physics}

\ifdefined\showcaptionsetup
 \PassOptionsToPackage{caption=false}{subfig}
\fi
\usepackage{subfig}
\makeatother

\usepackage{babel}
\begin{document}
\begin{frontmatter}
\title{An infinite grid of mesoscopic resistors – investigation and visualization
of ballistic conduction}
\author[rvt]{O. Urbański}
\ead{oliwier1459@gmail.com}
\address[rvt]{Faculty of Physics, Adam Mickiewicz University of Poznań, Uniwersytetu
Poznańskiego 2, 61-614 Poznań, Poland}
\begin{abstract}
A square lattice of mesoscopic resistors is considered. Each bond
is modeled as a narrow waveguide, while junctions are sources of elastic
scattering given by a scattering matrix $\mathbf{S}$. Symmetry and
unitarity constraints are used in a detailed way to simplify all possible
matrices $\mathbf{S}$. Energetic band structure of the system is
determined and visualized for exemplary parameters. Conductance between
external electrodes attached to two arbitrary nodes is given by the
Landauer formula. Thus the transmittance between the electrodes is
calculated. The rest of the paper is devoted to studying wave function
patterns arising from an electron injected through one electrode into
the system. It is observed that depending on its energy both dull
and localized or intricate and expansive patterns occur. It is justified
mathematically that fitting electron energy into an energy band can
be associated with emergence of complex patterns. Finally, possible
experimental realizations of the considered model are briefly mentioned.
\end{abstract}
\begin{keyword}
quantum interference effects \sep ballistic conduction \sep mesoscopic
systems
\end{keyword}
\end{frontmatter}

\section{Introduction}

Calculating the net resistance of a network of resistors is a classic
and practical school problem. Although larger structures generally
demand more computation, periodic and infinite lattices can be handled
analytically \citep{key-1,key-2}. In the case of the net resistance
measured between adjacent nodes (for uniform lattices), symmetry arguments
suffice to determine the answer \citep{key-1}. For alternating current,
resistance can be replaced by impedance, so that the mentioned method
applies to an even broader family of circuits.

\begin{figure*}
\centering
\subfloat[Square lattice\label{fig:Square-lattice}]{\centering{}\includegraphics[scale=0.65]{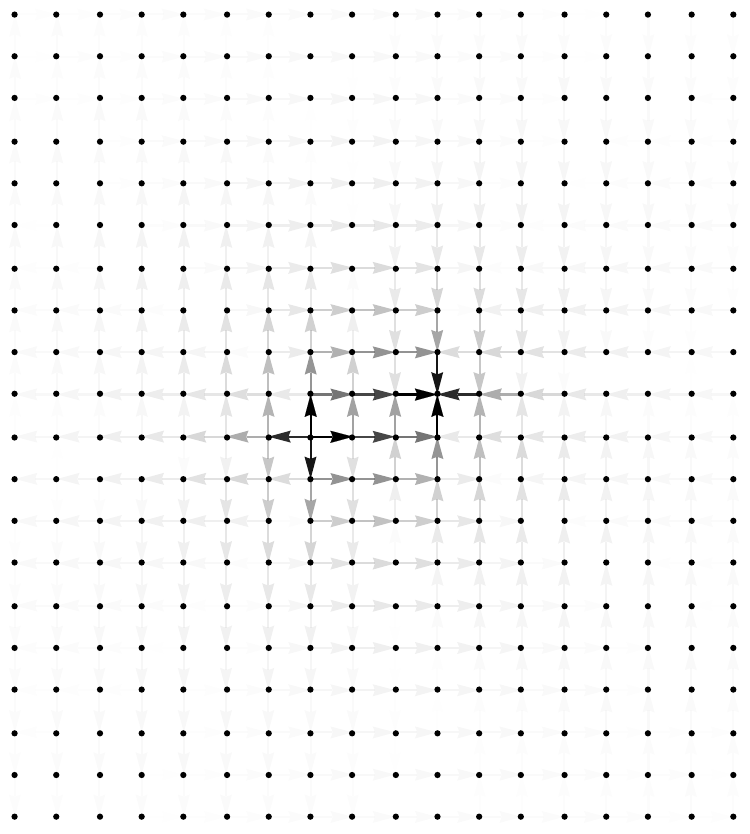}}\\\subfloat[Kagom\'e lattice\label{fig:Kagome-lattice}]{\centering{}\includegraphics[scale=0.65]{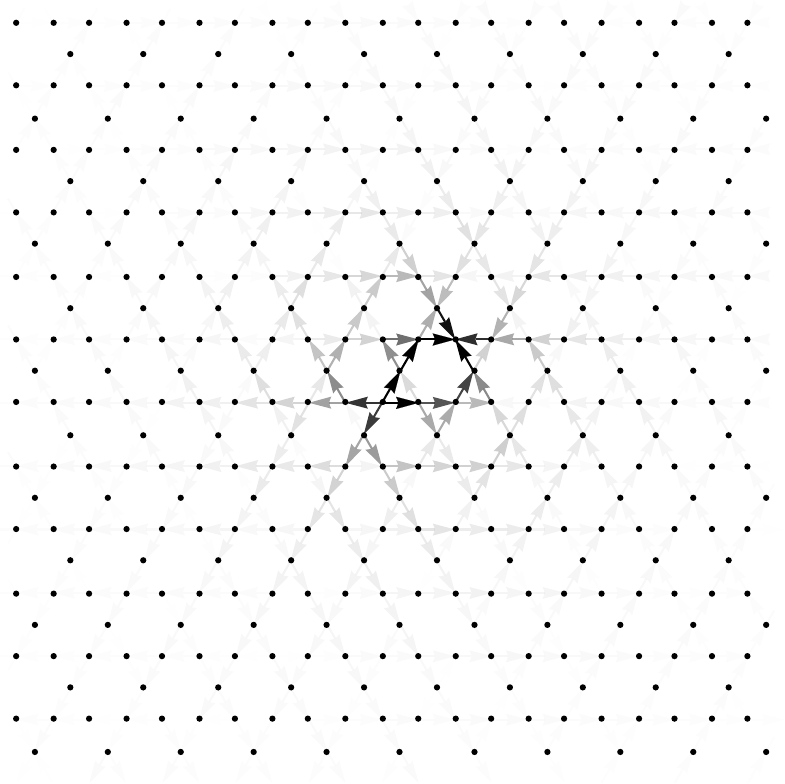}}\caption{Current distribution in infinite grids of classical resistors (arrows
indicate direction of the current and their opacity encode its value)\label{fig:Current-distribution-in}}
\end{figure*}

For arbitrary measuring nodes the problem is more difficult. However,
translational symmetry invites one to consider the discrete Fourier
transform of relevant quantities (switching to the so-called $k$-space).
Doing this within the framework of Green's functions yields the net
resistance \citep{key-2}. Current distribution can also be extracted
and visualized in a straightforward manner. Figure \ref{fig:Current-distribution-in}
shows such results for a square and Kagom\'e lattices obtained by
the author using an essentially identical approach to \citep{key-2}.

A fundamental assumption present to this point is the Ohm's law. However,
it ceases to hold for mesoscopic conductors (i. e. small enough that
the mean length of coherence is greater than the conduction path)
\citep{key-3,key-4,key-5,key-6,key-7}. Quantum (i. e. interference)
effects are of major importance, which modify calculations of the
net resistance. In particular, this quantity is no longer additive
for series connections \citep{key-4} and its inverse is not additive
for parallel connections \citep{key-6,key-8}. A natural question
arises: How do infinite grids of mesoscopic resistors work and are
they in any way similar to their classical counterpart?

This question has been partially answered in \citep{key-8} and \citep{key-9}.
The first work analyzes mesoscopic lattices of resistors to ultimately
facilitate quantum percolation, in which wires are randomly deleted.
Electrodes are attached to the opposite sides of a large rectangular
grid (they extend through an entire length of each side). This slightly
differs from the set-up considered in this work. In a full analogy
to the classical case (as in Fig. \ref{fig:Current-distribution-in}),
here electrodes are external wires connected to two different nodes.
The second mentioned paper focuses on a rectangular lattice with sides
$\ell_{1},\ell_{2}$ and investigates how the energy band structure
depends on $\ell_{1}/\ell_{2}$. Its title emphasizes that this system
is a natural generalization of the Kronig-Penney model introduced
in \citep{key-10} to two dimensions. Reference \citep{key-11} provides
a general and detailed treatment of finite counterparts of the considered
problem for arbitrary graphs.

This paper is structured as follows: In Sec. \ref{sec:Model} the
model for a grid of mesoscopic resistors is described. Sec. \ref{sec:Unitarity-and-symmetry}
shows how basic assumptions allow one to reduce the space of all possible
scattering matrices appearing in the model. Sec. \ref{sec:Band-structure}
presents a calculation of the band-structure, while Sec. \ref{sec:Transmittance-between-two}
tackles the original problem of resistance between the electrodes.
It is, however, Sec. \ref{sec:Wave-function-patterns} which constitutes
the heart of this paper, focusing on the features and visualization
of the current distribution (or equivalently the wave function pattern).
Possible experimental realizations of the considered system are discussed
in Sec. \ref{sec:Experimental-realization-and}, ended with a brief
summary of the entire work.

\section{Model\label{sec:Model}}

A mesoscopic resistor is modeled as a narrow waveguide, with its transverse
dimensions small enough, that the electron motion is essentially one-dimensional.
Therefore, only one mode (or channel) of propagation is assumed to
be available for the electrons. We consider a square lattice with
spacing $d$ and periodic boundary conditions. Intersections of the
wires form centers of scattering, each described by the same scattering
matrix $\mathbf{S}$. This formulation can account also for scattering
sources present in the wires themselves, provided that they are symmetric
and identical for every wire. The scattering from a half of every
conductor must be incorporated into the scattering matrix corresponding
to the closest junction.

To define $\mathbf{S}$, let us consider a single electron wave function
propagating in the system. Following \citep{key-8}, in a given wire,
it can be given as

\begin{equation}
\psi=Ae^{\mathrm{i}\kappa x}+Be^{-\mathrm{i}\kappa x},
\end{equation}
where $\mathrm{i}$ is the imaginary unit ($i$ is reserved for indexing
lattice sites), $A$ and $B$ are complex amplitudes, $\kappa$ is
the Fermi wave number and $x$ denotes position along the conductor
measured from its center. Let $\bm{\mathcal{I}}_{i}$ denote a vector
of complex amplitudes of waves incoming from each direction to site
$i$. Likewise, let $\bm{\mathcal{O}}_{i}$ denote the outgoing amplitudes:

\begin{equation}
\bm{\mathcal{I}}_{i}=\begin{pmatrix}\mathcal{I}_{1i}\\
\mathcal{I}_{2i}\\
\mathcal{I}_{3i}\\
\mathcal{I}_{4i}
\end{pmatrix},\;\bm{\mathcal{O}}_{i}=\begin{pmatrix}\mathcal{O}_{1i}\\
\mathcal{O}_{2i}\\
\mathcal{O}_{3i}\\
\mathcal{O}_{4i}
\end{pmatrix}.\label{IO}
\end{equation}
Numbers $1,2,3,4$ correspond to four different directions according
to the following diagram:

\begin{equation}
\begin{array}{cccccc}
 &  & \mathcal{I}_{4i} & \mathcal{O}_{4i}\\
 &  & \downarrow & \uparrow\\
\mathcal{O}_{1i} & \leftarrow &  &  & \leftarrow & \mathcal{I}_{3i}\\
\mathcal{I}_{1i} & \rightarrow &  &  & \rightarrow & \mathcal{O}_{3i}\\
 &  & \downarrow & \uparrow\\
 &  & \mathcal{O}_{2i} & \mathcal{I}_{2i}
\end{array}.\label{diagram}
\end{equation}
Finally, $\mathbf{S}$ being a $4\times4$ matrix is defined via a
relation
\begin{equation}
\bm{\mathcal{O}}_{i}=\mathbf{S}\bm{\mathcal{I}}_{i}.\label{O=00003DSI}
\end{equation}

In order to measure the net resistance between two arbitrary nodes
$i_{1}$ and $i_{2}$, external electrodes are needed. They inevitably
change the scattering at $i_{1},i_{2}$, so that two new $5\times5$
scattering matrices $\mathbf{S}_{1}$ and $\mathbf{S}_{2}$ are needed
there. Treating the electrodes as two pins of the entire system, it
can be described by a $2\times2$ net scattering matrix
\begin{equation}
\mathbf{S}_{\mathrm{net}}=\begin{pmatrix}r & t\\
t & r
\end{pmatrix}.
\end{equation}
Both electrodes are on equal footing and thus the reflection and transmission
amplitudes $r,t$ are identical for both ways of propagation.

The net conductance $G$ between the electrodes is given by the Landauer
formula \citep{key-3,key-5}. We use its version including the Sharvin
contact resistance \citep{key-7}:

\begin{equation}
G=\frac{e^{2}}{h}\left|t\right|^{2}.
\end{equation}

\section{Unitarity and symmetry constraints on the scattering matrix\label{sec:Unitarity-and-symmetry}}

\citep{key-6} gives a full $3\times3$ scattering matrix for a one-to-two
symmetric junction (which by assumption does not backscatter). Its
form is almost entirely determined by unitarity and spatial symmetry
of the system. We want to use such arguments to restrict both $\mathbf{S}$
and $\mathbf{S}_{1}$ ($\mathbf{S}_{2}=\mathbf{S}_{1}$). Starting
with $\mathbf{S}$, a 4-fold rotary symmetry of each junction, together
with an inversion center, enforces the following form:

\begin{equation}
\mathbf{S}=\begin{pmatrix}b & s & f & s\\
s & b & s & f\\
f & s & b & s\\
s & f & s & b
\end{pmatrix}e^{\mathrm{i}\kappa d}.
\end{equation}
Factor $e^{\mathrm{i}\kappa d}$ is customary and reflects the phase
acquired by traveling the lattice constant.

Elements of matrix $\mathbf{S}$ depend only on the difference of
their indices. For this reason, $\mathbf{S}$ can be diagonalized
by the discrete Fourier transform. Then $\mathbf{S}=\mathbf{U}^{\dagger}\mathbf{D}\mathbf{U}$,
where $\mathbf{U}$ is a unitary matrix representing the discrete
Fourier transform and $\mathbf{D}$ is a complex diagonal matrix.
Unitarity is now equivalent to:

\begin{align}
\mathbf{1} & =\mathbf{S}\mathbf{S}^{\dagger}=\mathbf{U}^{\dagger}\mathbf{D}\mathbf{D}^{\dagger}\mathbf{U}=\mathbf{U}^{\dagger}\mathbf{D}\mathbf{D}^{*}\mathbf{U}\nonumber \\
 & \Leftrightarrow\mathbf{D}\mathbf{D}^{*}=\mathbf{1}.
\end{align}

Thus all eigenvalues of $\mathbf{S}$ must be of unit module, which
can be written as:

\begin{equation}
\begin{cases}
b+f+2s & =e^{\mathrm{i}\phi_{1}}\\
b-f & =e^{\mathrm{i}\phi_{2}}\\
b+f-2s & =e^{\mathrm{i}\phi_{3}}
\end{cases}.\label{bfs}
\end{equation}
Solving for $b,s,f$, we get:
\begin{equation}
\begin{cases}
b=\frac{1}{4}\left(e^{\mathrm{i}\phi_{1}}+e^{\mathrm{i}\phi_{3}}\right)+\frac{1}{2}e^{\mathrm{i}\phi_{2}}\\
s=\frac{1}{4}\left(e^{\mathrm{i}\phi_{1}}-e^{\mathrm{i}\phi_{3}}\right)\\
f=\frac{1}{4}\left(e^{\mathrm{i}\phi_{1}}+e^{\mathrm{i}\phi_{3}}\right)-\frac{1}{2}e^{\mathrm{i}\phi_{2}}
\end{cases}.\label{b,s,f}
\end{equation}

It is interesting that thanks to high symmetry of the scattering potential
and unitarity of $\mathbf{S}$, we managed to reduce $16$ complex
matrix elements down to just $3$ real phases. Equation \eqref{b,s,f}
implies an interesting inequality, namely $\left|s\right|\leq\frac{1}{2}$,
so $\left|s\right|^{2}\leq\frac{1}{4}$. This means that no more than
$25\%$ of electrons get scattered to the right (or left). Taking
$e^{\mathrm{i}\phi_{2}}$ out of the matrix and introducing $\varphi_{1}=\phi_{1}-\phi_{2}$,
$\varphi_{3}=\phi_{3}-\phi_{2}$, we obtain:

\begin{equation}
\mathbf{S}=\frac{e^{\mathrm{i}\left(\kappa d+\phi_{2}\right)}}{4}\begin{pmatrix}q_{+}+2 & q_{-} & q_{+}-2 & q_{-}\\
q_{-} & q_{+}+2 & q_{-} & q_{+}-2\\
q_{+}-2 & q_{-} & q_{+}+2 & q_{-}\\
q_{-} & q_{+}-2 & q_{-} & q_{+}+2
\end{pmatrix},\label{S}
\end{equation}
where $q_{+}=e^{\mathrm{i}\varphi_{1}}+e^{\mathrm{i}\varphi_{3}}$,
$q_{-}=e^{\mathrm{i}\varphi_{1}}-e^{\mathrm{i}\varphi_{3}}$. The
scattering matrix implicitly used in \citep{key-8} and \citep{key-9}
(choice referred to as $\delta$-coupling with $\alpha=0$) is obtained
by substituting $\varphi_{1}=\pi,\,\varphi_{3}=0,\,\phi_{2}=\pi$.
In general, the introduced phases may depend on $\kappa$. However,
for sufficiently small connections between mesoscopic conductors with
comparison to $2\pi/\kappa$, we can expect the phases to be essentially
independent of $\kappa$ (which is assumed throughout the rest of
the paper).

Scattering matrix $\mathbf{S}_{1}$ can be written as a block matrix

\begin{equation}
\mathbf{S}_{1}=\begin{pmatrix}\mathbf{S}_{1}^{\mathrm{red}} & \mathbf{v}_{1}\\
\mathbf{w}_{1} & \left(\mathbf{S}_{1}\right)_{55}
\end{pmatrix},\label{S1}
\end{equation}
where $\mathbf{S}_{1}^{\mathrm{red}}$ has dimensions $4\times4$,
$\mathbf{w}_{1}$ is $1\times4$ and $\mathbf{v}_{1}$ is $4\times1$
(superscript ``$\mathrm{red}$'' stands for ``reduced dimensions'').
The fifth index corresponds to the external electrode, which is perpendicular
to the lattice. Assuming (similarly to \citep{key-6}) no backscattering
from the electrodes, we can take $\mathbf{v}_{1}^{T}=\mathbf{w}_{1}=\begin{pmatrix}\frac{1}{2} & \frac{1}{2} & \frac{1}{2} & \frac{1}{2}\end{pmatrix}$
and $\left(\mathbf{S}_{1}\right)_{55}=0$. Additionally, full symmetry
of the remaining four wires leads to
\begin{equation}
\mathbf{S}_{1}^{\mathrm{red}}=\begin{pmatrix}a_{0} & a_{1} & a_{2} & a_{1}\\
a_{1} & a_{0} & a_{1} & a_{2}\\
a_{2} & a_{1} & a_{0} & a_{1}\\
a_{1} & a_{2} & a_{1} & a_{0}
\end{pmatrix}.
\end{equation}

Again, matrix elements of $\mathbf{S}_{1}^{\mathrm{red}}$ depend
only on the difference of indices. This invites us to use the discrete
Fourier transform. Thus, let

\begin{equation}
\mathbf{F}=\begin{pmatrix}\mathbf{F}^{\mathrm{red}} & 0\\
0 & 1
\end{pmatrix},
\end{equation}
where $\mathbf{F}^{\mathrm{red}}$ is a $4\times4$ matrix with elements
given by:

\begin{equation}
\left(\mathbf{F}^{\mathrm{red}}\right)_{nm}=\frac{1}{2}e^{-2\pi\mathrm{i}\frac{\left(n-1\right)\left(m-1\right)}{4}},
\end{equation}
where $n,m=1,2,3,4$. $\mathbf{F}$ is a unitary matrix, which satisfies

\begin{equation}
\mathbf{F}\mathbf{S}_{1}\mathbf{F}^{\dagger}=\begin{pmatrix}2\tilde{a}_{0} & 0 & 0 & 0 & 1\\
0 & 2\tilde{a}_{1} & 0 & 0 & 0\\
0 & 0 & 2\tilde{a}_{2} & 0 & 0\\
0 & 0 & 0 & 2\tilde{a}_{1} & 0\\
1 & 0 & 0 & 0 & 0
\end{pmatrix}.
\end{equation}

Sequence $\left(\tilde{a}_{0},\tilde{a}_{1},\tilde{a}_{2},\tilde{a}_{1}\right)$
is a discrete Fourier transform of sequence $\left(a_{0},a_{1},a_{2},a_{1}\right)$.
$\mathbf{F}$ and $\mathbf{S}_{1}$ are unitary, so $\mathbf{F}\mathbf{S}_{1}\mathbf{F}^{\dagger}$
is as well. It implies $\tilde{a}_{0}=0$. $2\tilde{a}_{1}$ and $2\tilde{a}_{2}$
are eigenvalues of $\mathbf{F}\mathbf{S}_{1}\mathbf{F}^{\dagger}$,
so they must also be the eigenvalues of $\mathbf{S}_{1}$. From this
it follows, that they are of unit module:

\begin{equation}
2\tilde{a}_{1}=e^{\mathrm{i}f_{1}},\:2\tilde{a}_{2}=e^{\mathrm{i}f_{2}}.
\end{equation}

Performing the inverse discrete Fourier transform, we get:

\begin{equation}
\begin{cases}
a_{0}=\frac{1}{2}e^{\mathrm{i}f_{2}}\left(\frac{1}{2}+e^{\mathrm{i}f}\right)\\
a_{1}=-\frac{1}{4}e^{\mathrm{i}f_{2}}\\
a_{2}=\frac{1}{2}e^{\mathrm{i}f_{2}}\left(\frac{1}{2}-e^{\mathrm{i}f}\right)
\end{cases},
\end{equation}
with $f=f_{1}-f_{2}$. Finally:

\begin{equation}
\mathbf{S}_{1}^{\mathrm{red}}=\frac{e^{\mathrm{i}f_{2}}}{2}\begin{pmatrix}\frac{1}{2}+e^{\mathrm{i}f} & -\frac{1}{2} & \frac{1}{2}-e^{\mathrm{i}f} & -\frac{1}{2}\\
-\frac{1}{2} & \frac{1}{2}+e^{\mathrm{i}f} & -\frac{1}{2} & \frac{1}{2}-e^{\mathrm{i}f}\\
\frac{1}{2}-e^{\mathrm{i}f} & -\frac{1}{2} & \frac{1}{2}+e^{\mathrm{i}f} & -\frac{1}{2}\\
-\frac{1}{2} & \frac{1}{2}-e^{\mathrm{i}f} & -\frac{1}{2} & \frac{1}{2}+e^{\mathrm{i}f}
\end{pmatrix},\label{S1red}
\end{equation}
which contains only two real phases.

\section{Band structure\label{sec:Band-structure}}

The kinetic energy of an electron propagating in a waveguide can be
written as $E=\hslash^{2}\kappa^{2}/2m+E_{0}$. $m$ stands for the
effective mass of an electron and $E_{0}$ is the energy associated
with transverse behavior of the wave function. Since only one mode
is considered, we can fix the energy scale by setting $E_{0}=0$.
Therefore, determining the band structure (in the absence of external
electrodes) requires finding such $\kappa$ that there exist amplitudes
from Eq. \eqref{IO}, which satisfy Eq. \eqref{O=00003DSI}. One more
geometric condition is needed, namely, an outgoing amplitude is simultaneously
an incoming amplitude for a neighboring site:

\begin{equation}
\begin{pmatrix}\mathcal{I}_{1i}\\
\mathcal{I}_{2i}\\
\mathcal{I}_{3i}\\
\mathcal{I}_{4i}
\end{pmatrix}=\begin{pmatrix}\mathcal{O}_{3\left(i+\left(-1,0\right)\right)}\\
\mathcal{O}_{4\left(i+\left(0,-1\right)\right)}\\
\mathcal{O}_{1\left(i+\left(1,0\right)\right)}\\
\mathcal{O}_{2\left(i+\left(0,1\right)\right)}
\end{pmatrix},\label{I in terms of O}
\end{equation}
where $i$ is a pair of coordinates labeling a corresponding site.
Relations \eqref{O=00003DSI} and \eqref{I in terms of O} form a
gigantic set of linear equations. Similarly to the classical case,
switching to the $k$-space may be beneficial, because translationally
invariant objects simplify. Thus, let $\bm{\mathcal{I}}_{k}$ and
$\bm{\mathcal{O}}_{k}$ represent discrete Fourier transforms of $\bm{\mathcal{I}}_{i}$
and $\bm{\mathcal{O}}_{i}$ respectively, given by:

\begin{equation}
\bm{\mathcal{I}}_{k}=\frac{1}{\sqrt{N}}\sum_{i}\bm{\mathcal{I}}_{i}e^{-\mathrm{i}k\cdot i},
\end{equation}

\begin{equation}
\bm{\mathcal{O}}_{k}=\frac{1}{\sqrt{N}}\sum_{i}\bm{\mathcal{O}}_{i}e^{-\mathrm{i}k\cdot i},
\end{equation}
with $N$ representing the total number of sites. Now, both Eqs. \eqref{O=00003DSI}
and \eqref{I in terms of O} take simpler on-site forms (i. e. such
that involve only one index of a $k$-space lattice site):
\begin{equation}
\bm{\mathcal{O}}_{k}=\mathbf{S}\bm{\mathcal{I}}_{k},\label{Ok=00003DSIk}
\end{equation}

\begin{equation}
\begin{pmatrix}\mathcal{I}_{1k}\\
\mathcal{I}_{2k}\\
\mathcal{I}_{3k}\\
\mathcal{I}_{4k}
\end{pmatrix}=\begin{pmatrix}\mathcal{O}_{3k}e^{-\mathrm{i}k_{x}}\\
\mathcal{O}_{4k}e^{-\mathrm{i}k_{y}}\\
\mathcal{O}_{1k}e^{\mathrm{i}k_{x}}\\
\mathcal{O}_{2k}e^{\mathrm{i}k_{y}}
\end{pmatrix}.\label{I in terms of O-1}
\end{equation}

Equation \eqref{I in terms of O-1} can be written in a matrix form:

\begin{equation}
\bm{\mathcal{I}}_{k}=\mathbf{M}_{k}\bm{\mathcal{O}}_{k},\label{Ik=00003DMOk}
\end{equation}
where
\begin{equation}
\mathbf{M}_{k}=\begin{pmatrix}0 & 0 & e^{-\mathrm{i}k_{x}} & 0\\
0 & 0 & 0 & e^{-\mathrm{i}k_{y}}\\
e^{\mathrm{i}k_{x}} & 0 & 0 & 0\\
0 & e^{\mathrm{i}k_{y}} & 0 & 0
\end{pmatrix}.\label{M}
\end{equation}

Joining Eqs. \eqref{Ok=00003DSIk} and \eqref{Ik=00003DMOk}, we get:

\begin{equation}
\bm{\mathcal{O}}_{k}=\mathbf{S}\mathbf{M}_{k}\bm{\mathcal{O}}_{k}.\label{smok}
\end{equation}

Equality \eqref{smok} means that for every $k$, $\bm{\mathcal{O}}_{k}$
is an eigenvector of $\mathbf{S}\mathbf{M}_{k}$ with eigenvalue $1$
(or just a zero vector). Matrix $\mathbf{M}_{k}$ is both Hermitian
and unitary. Scattering matrix is also unitary, so $\mathbf{S}\mathbf{M}_{k}$
is as well. Thus its eigenvalues are of the form $e^{\mathrm{i}\phi}$,
for some real number $\phi$. If for at least one $k=k_{0}$ $\mathbf{S}\mathbf{M}_{k}$
has at least one eigenvalue equal to $1$ (with eigenvector $\bm{\mathcal{O}}_{k_{0}}$),
then $\mathcal{O}_{i}$ being an inverse Fourier transform of $\bm{\mathcal{O}}_{k_{0}}\delta_{kk_{0}}$
satisfies the initial problem. Therefore, in order to find the spectrum,
for every $k$ (which is non-dimensional quasi-momentum), $\kappa$
(being the microscopic wave number) should be adjusted so as to match
$e^{\mathrm{i}\phi}=1$. This gives four branches of dispersion relation
(as there are four eigenvalues) $E_{n}\left(k\right)=\hslash^{2}\kappa_{n}^{2}\left(k\right)/2m$
($n$ indexes branches). These branches, as it turns out, will have
subbranches.

Matrix multiplication $\mathbf{S}\mathbf{M}_{k}$ gives:

\begin{equation}
\mathbf{S}\mathbf{M}_{k}=e^{\mathrm{i}\left(\kappa d+\phi_{2}\right)}\mathbf{m}_{k},\label{SMk}
\end{equation}
where:

\begin{equation}
\mathbf{m}_{k}=\frac{1}{4}\begin{pmatrix}e^{\mathrm{i}k_{x}}\left(q_{+}-2\right) & e^{\mathrm{i}k_{y}}q_{-} & e^{-\mathrm{i}k_{x}}\left(q_{+}+2\right) & e^{-\mathrm{i}k_{y}}q_{-}\\
e^{\mathrm{i}k_{x}}q_{-} & e^{\mathrm{i}k_{y}}\left(q_{+}-2\right) & e^{-\mathrm{i}k_{x}}q_{-} & e^{-\mathrm{i}k_{y}}\left(q_{+}+2\right)\\
e^{\mathrm{i}k_{x}}\left(q_{+}+2\right) & e^{\mathrm{i}k_{y}}q_{-} & e^{-\mathrm{i}k_{x}}\left(q_{+}-2\right) & e^{-\mathrm{i}k_{y}}q_{-}\\
e^{\mathrm{i}k_{x}}q_{-} & e^{\mathrm{i}k_{y}}\left(q_{+}+2\right) & e^{-\mathrm{i}k_{x}}q_{-} & e^{-\mathrm{i}k_{y}}\left(q_{+}-2\right)
\end{pmatrix}.\label{m}
\end{equation}

All four eigenvalues of $\mathbf{m}_{k}$ are of unit modulus. We
can write them as $e^{\mathrm{i}u_{n}}$, where $n$ indexes four
real phases. To force the $n$-th eigenvalue of $\mathbf{S}\mathbf{M}_{k}$
to be $1$ (according to Eq. \eqref{SMk}), we have to put:

\begin{equation}
\kappa_{nl}d+\phi_{2}=-u_{n}+2\pi l.\label{kappa}
\end{equation}
$l$ is an integer which indexes the mentioned subbranches. Equation
\eqref{kappa} can be written as:

\begin{equation}
\kappa_{nl}=\frac{-u_{n}-\phi_{2}+2\pi l}{d}.\label{kappa again}
\end{equation}

\begin{figure}
\centering{}\includegraphics[scale=0.45]{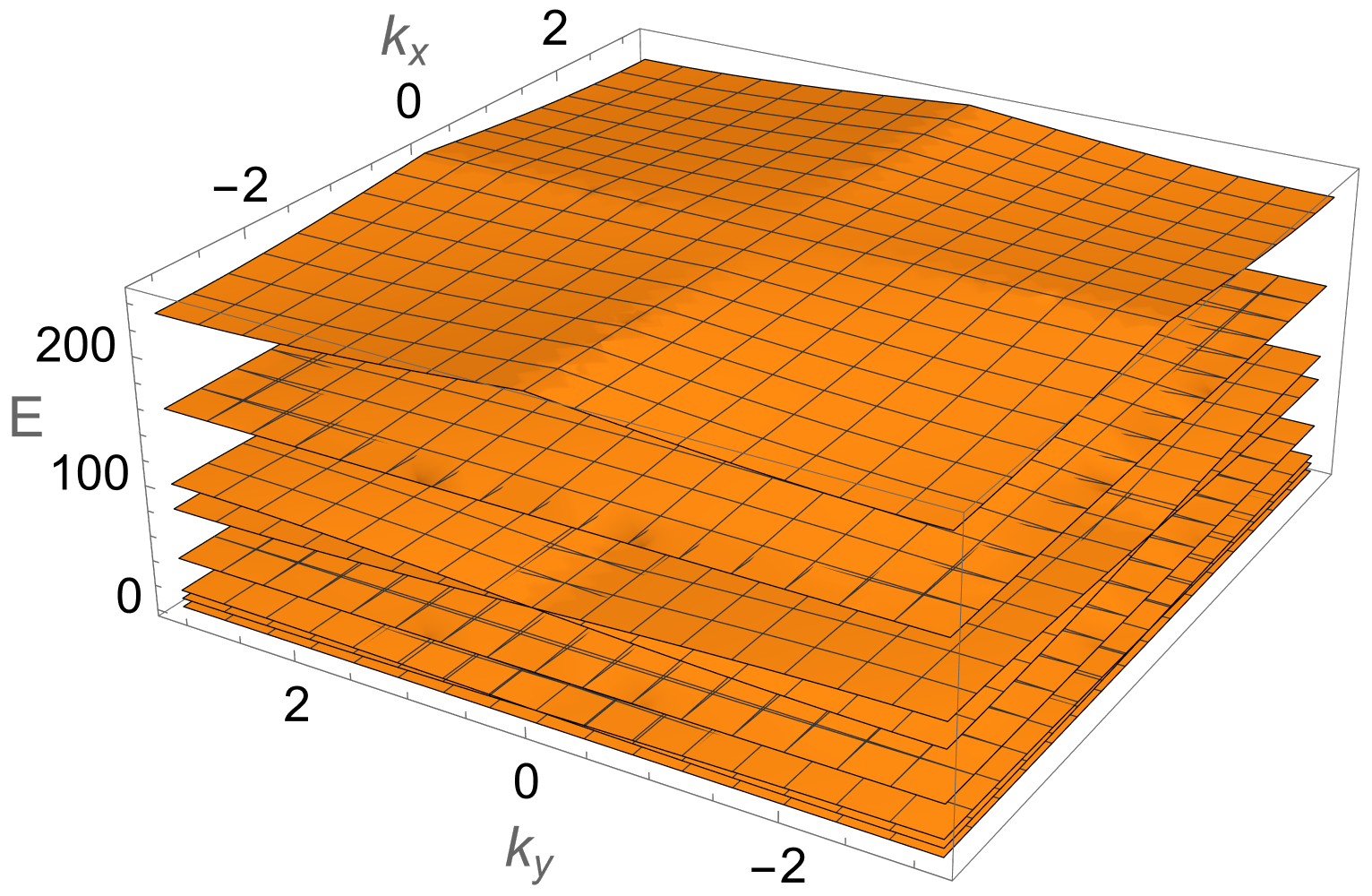}\caption{Band structure\label{fig:Band-structure}}
\end{figure}

\begin{figure*}
\centering{}\subfloat[The lowest band]{\centering{}\includegraphics[scale=0.45]{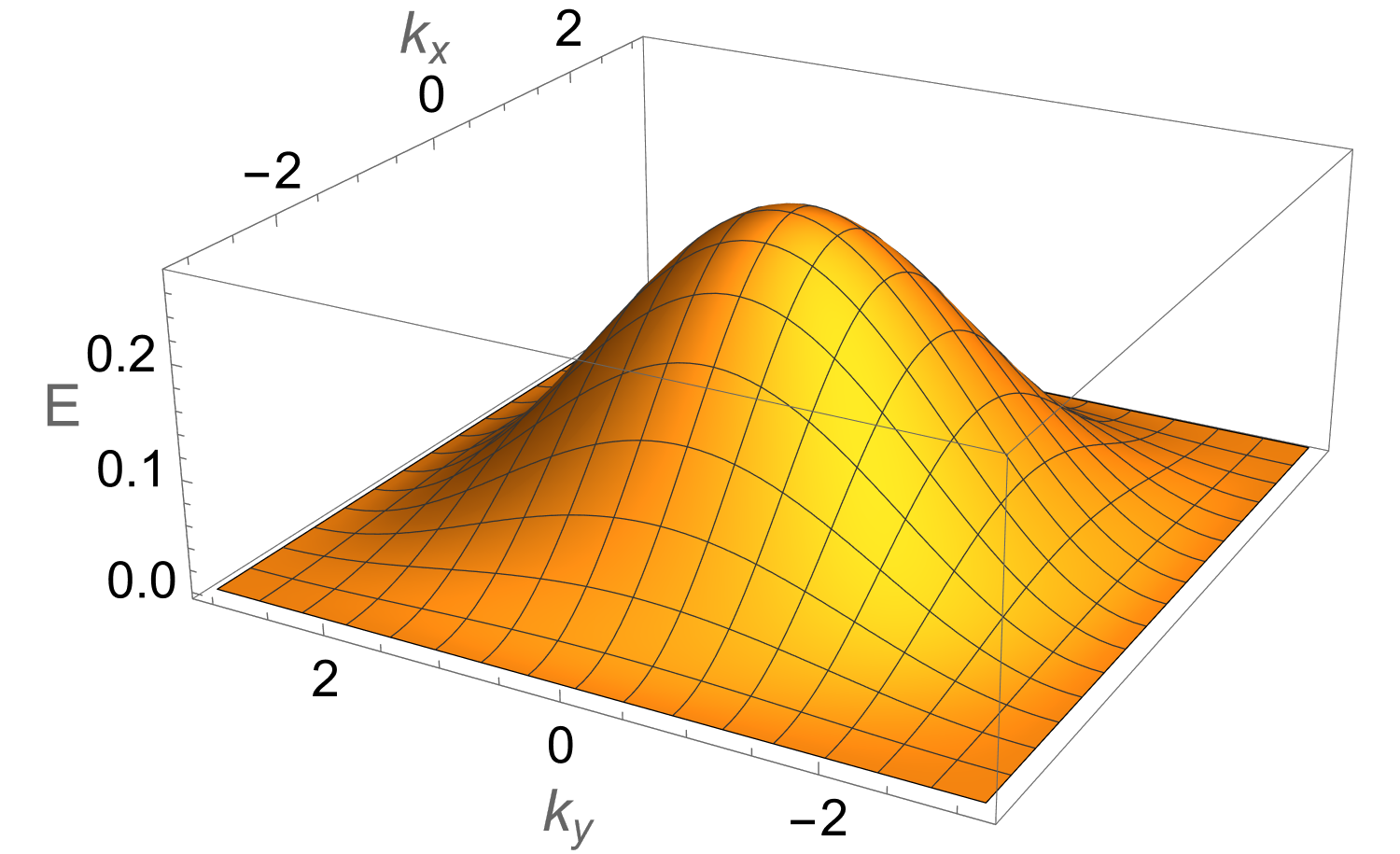}}\\\subfloat[The two next bands]{\centering{}\includegraphics[scale=0.45]{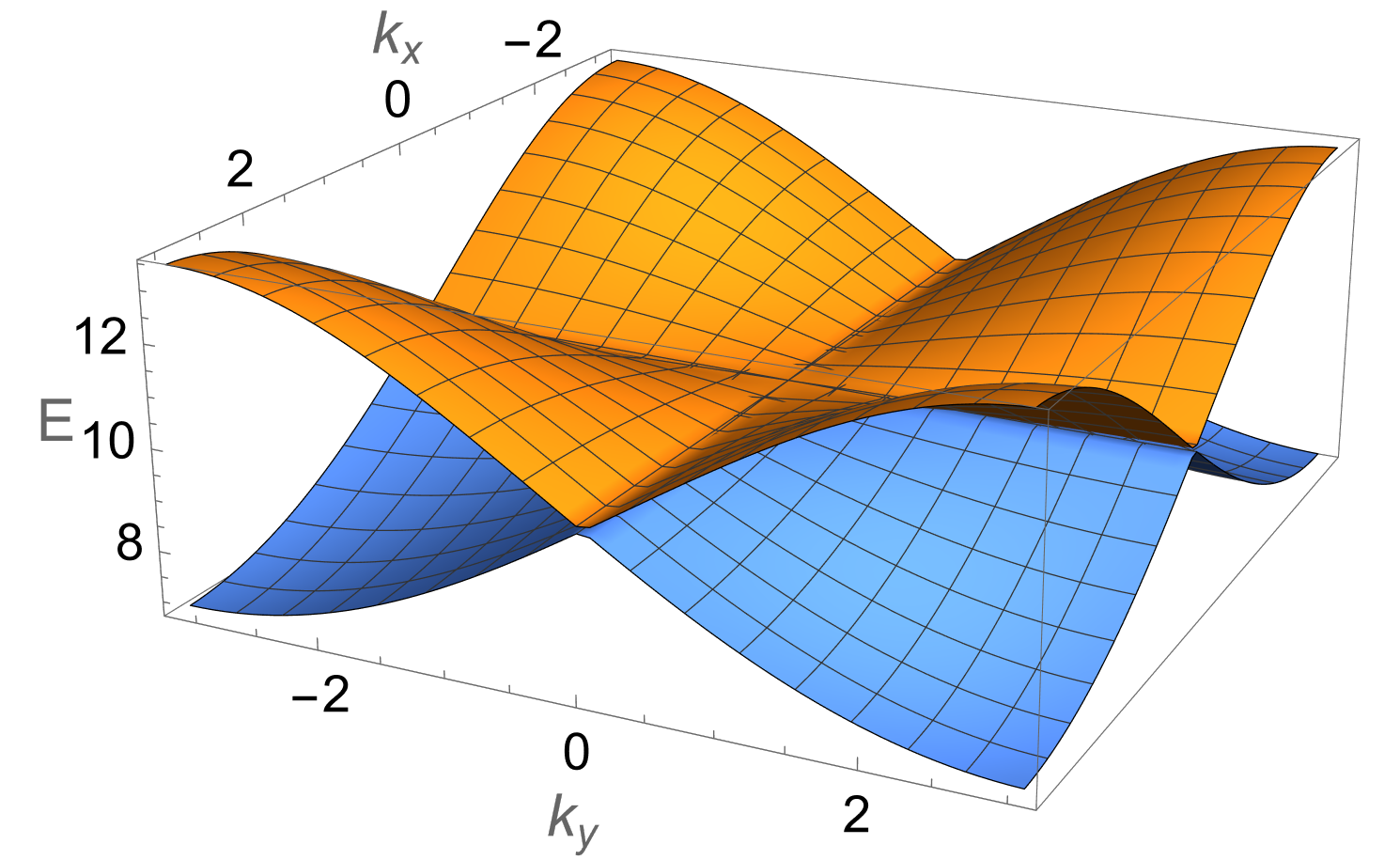}}\caption{The lowest three bands\label{fig:The-lowest-three}}
\end{figure*}

Calculating $u_{n}$ reduces to finding roots of the fourth order
characteristic polynomial of matrix $\mathbf{m}_{k}$. This can be
done by means of the Ferrari method \citep{key-12} (which result
takes too much space to be written explicitly here) or numerically.
However, a simple analytical formula can be found for $\varphi_{3}=-\varphi_{1}$.
Physically this condition means, that $b$ and $f$ are are shifted
in phase by $\pi$, while $s$ is out of phase by $\pi/2$ (relative
to $b$ or $f$). \ref{sec:Derivation-of-the} gives a derivation
of the following formula for the dispersion relation:

\begin{equation}
E=\frac{\hslash^{2}}{2m}\left[\frac{\pm_{1}\arccos\left(\frac{\sigma}{2}\pm_{2}\sqrt{\frac{\sigma^{2}}{4}-p}\right)-\phi_{2}+2\pi l}{d}\right]^{2},\label{E}
\end{equation}
where

\begin{equation}
\sigma=-\frac{1}{2}\left(1-\cos\varphi_{1}\right)\left(\cos k_{x}+\cos k_{y}\right),
\end{equation}

\begin{equation}
p=\frac{1}{2}\left[\cos k_{x}\cos k_{y}\left(1-\cos\varphi_{1}\right)-\cos\varphi_{1}-1\right],
\end{equation}
$\pm_{1}$ and $\pm_{2}$ are independent plus-minus signs generating
four branches and $l$ is an arbitrary integer giving rise to further
subbranches. Figures \ref{fig:Band-structure} and \ref{fig:The-lowest-three}
show the dispersion relation plots following from Eq. \eqref{E} for
$\frac{\hslash^{2}}{2m}=1$, $d=1$, $\phi_{2}=0$ and $\varphi_{1}=\pi/6$.
It is important to note, that the presence of gaps depends on the
chosen parameters. For example, $\varphi_{1}=\pi,\,\varphi_{3}=0,\,\phi_{2}=\pi$,
corresponding to \citep{key-8}, lead to a gapless spectrum.

\section{Transmittance between two electrodes\label{sec:Transmittance-between-two}}

In the previous situation (i. e. without the electrodes), the scattering
relation Eq. \eqref{O=00003DSI} was the same for every site $i$.
Here we introduce two electrodes at sites $i_{1}$ and $i_{2}$, so
that Eq. \eqref{O=00003DSI} has to be changed at these sites. Let
$\mathcal{I}_{1}^{\mathrm{el}}$ and $\mathcal{O}_{1}^{\mathrm{el}}$
denote the incoming and outgoing complex amplitude respectively at
the first electrode. $\mathcal{I}_{2}^{\mathrm{el}}$ and $\mathcal{O}_{2}^{\mathrm{el}}$
are defined similarly for the second electrode. The scattering relation
takes then the following form:

\begin{equation}
\begin{cases}
\begin{pmatrix}\bm{\mathcal{O}}_{i_{1}}\\
\mathcal{O}_{1}^{\mathrm{el}}
\end{pmatrix}=\mathbf{S}_{1}\begin{pmatrix}\bm{\mathcal{I}}_{i_{1}}\\
\mathcal{I}_{1}^{\mathrm{el}}
\end{pmatrix}, & \text{first electrode}\\
\begin{pmatrix}\bm{\mathcal{O}}_{i_{2}}\\
\mathcal{O}_{2}^{\mathrm{el}}
\end{pmatrix}=\mathbf{S}_{2}\begin{pmatrix}\bm{\mathcal{I}}_{i_{2}}\\
\mathcal{I}_{2}^{\mathrm{el}}
\end{pmatrix}, & \text{second electrode}\\
\bm{\mathcal{O}}_{i}=\mathbf{S}\bm{\mathcal{I}}_{i}, & i\neq i_{1},i_{2}
\end{cases}.\label{O=00003DSI extended}
\end{equation}

Equation \eqref{I in terms of O} remains unchanged and so does Eq.
\eqref{Ik=00003DMOk}.

The goal is to determine the ``net scattering matrix'' $\mathbf{S}_{\mathrm{net}}$,
which relates $\left(\mathcal{I}_{1}^{\mathrm{el}},\mathcal{I}_{2}^{\mathrm{el}}\right)$
to $\left(\mathcal{O}_{1}^{\mathrm{el}},\mathcal{O}_{2}^{\mathrm{el}}\right)$,
by

\begin{equation}
\begin{pmatrix}\mathcal{O}_{1}^{\mathrm{el}}\\
\mathcal{O}_{2}^{\mathrm{el}}
\end{pmatrix}=\mathbf{S}_{\mathrm{net}}\begin{pmatrix}\mathcal{I}_{1}^{\mathrm{el}}\\
\mathcal{I}_{2}^{\mathrm{el}}
\end{pmatrix},\label{Snet}
\end{equation}
together with the field of amplitudes $\bm{\mathcal{I}}_{i}$ and
$\bm{\mathcal{O}}_{i}$. Keeping only the first four components in
the first two equations in Eq. \eqref{O=00003DSI extended}, we get
(using notation introduced in Eq. \eqref{S1}):

\begin{equation}
\begin{cases}
\bm{\mathcal{O}}_{i_{1}}=\mathbf{S}_{1}^{\mathrm{red}}\bm{\mathcal{I}}_{i_{1}}+\mathcal{I}_{1}^{\mathrm{el}}\mathbf{v}_{1} & \text{first electrode}\\
\bm{\mathcal{O}}_{i_{2}}=\mathbf{S}_{2}^{\mathrm{red}}\bm{\mathcal{I}}_{i_{2}}+\mathcal{I}_{2}^{\mathrm{el}}\mathbf{v}_{2} & \text{second electrode}\\
\bm{\mathcal{O}}_{i}=\mathbf{S}\bm{\mathcal{I}}_{i} & i\neq i_{1},i_{2}
\end{cases}.\label{O=00003DSI reduced}
\end{equation}
Although the full translational symmetry has been spoiled by introducing
the electrodes, performing the discrete Fourier transform is still
promising. The contribution from sites $i_{1}$ and $i_{2}$ will
break out from the previously encountered regularity, so it has to
be handled with special care. Amplitude field $\bm{\mathcal{O}}_{i}$
given by Eq. \eqref{O=00003DSI reduced} attains the following form
in the $k$-space:

\begin{align}
\bm{\mathcal{O}}_{k} & =\mathbf{S}\bm{\mathcal{I}}_{k}+\frac{\left(\mathbf{S}_{1}^{\mathrm{red}}\bm{\mathcal{I}}_{i_{1}}+\mathcal{I}_{1}^{\mathrm{el}}\mathbf{v}_{1}-\mathbf{S}\bm{\mathcal{I}}_{i_{1}}\right)e^{-\mathrm{i}k\cdot i_{1}}}{\sqrt{N}}\nonumber \\
 & +\frac{\left(\mathbf{S}_{2}^{\mathrm{red}}\bm{\mathcal{I}}_{i_{2}}+\mathcal{I}_{2}^{\mathrm{el}}\mathbf{v}_{2}-\mathbf{S}\bm{\mathcal{I}}_{i_{2}}\right)e^{-\mathrm{i}k\cdot i_{2}}}{\sqrt{N}}.\label{Ok}
\end{align}

Matrix $\mathbf{M}_{k}$ given by Eq. \eqref{M} is simultaneously
Hermitian and unitary and thus it is its own inverse. Thus Eq. \eqref{Ik=00003DMOk}
can be transformed into

\begin{equation}
\bm{\mathcal{O}}_{k}=\mathbf{M}_{k}\bm{\mathcal{I}}_{k}.\label{Ok=00003DMIk}
\end{equation}
Without loss of generality we can take $i_{1}=\left(0,0\right)$.
Substituting Eq. \eqref{Ok=00003DMIk} to \eqref{Ok} and solving
for $\bm{\mathcal{I}}_{k}$, we get:

\begin{align}
\bm{\mathcal{I}}_{k} & =\frac{1}{\sqrt{N}}\left\{ \left(\mathbf{M}_{k}-\mathbf{S}\right)^{-1}\left[\left(\mathbf{S}_{1}^{\mathrm{red}}-\mathbf{S}\right)\bm{\mathcal{I}}_{i=0}+\mathcal{I}_{1}^{\mathrm{el}}\mathbf{v}_{1}\right]\right.\nonumber \\
 & \left.+\left(\mathbf{M}_{k}-\mathbf{S}\right)^{-1}e^{-\mathrm{i}k\cdot i_{2}}\left[\left(\mathbf{S}_{2}^{\mathrm{red}}-\mathbf{S}\right)\bm{\mathcal{I}}_{i_{2}}+\mathcal{I}_{2}^{\mathrm{el}}\mathbf{v}_{2}\right]\right\} .\label{Ik}
\end{align}

The problem with Eq. \eqref{Ik} is that the incoming amplitudes we
want to solve for are present also on the right-hand-side. According
to the inverse discrete Fourier transform, $\bm{\mathcal{I}}_{i=0}=N^{-1/2}\sum_{k}\bm{\mathcal{I}}_{k}$
and $\bm{\mathcal{I}}_{i_{2}}=N^{-1/2}\sum_{k}\bm{\mathcal{I}}_{k}e^{\mathrm{i}k\cdot i_{2}}$.
Acting on Eq. \eqref{Ik} first with $N^{-1/2}\sum_{k}\cdot$ and
then with $N^{-1/2}\sum_{k}\cdot\,e^{\mathrm{i}k\cdot i_{2}}$ ($\cdot$
represents the argument of an operator) we obtain a set of two linear
equations for $\bm{\mathcal{I}}_{i=0}$ and $\bm{\mathcal{I}}_{i_{2}}$.
Let:

\begin{equation}
\mathbf{R}_{i}=\frac{1}{N}\sum_{k}\left(\mathbf{M}_{k}-\mathbf{S}\right)^{-1}e^{\mathrm{i}k\cdot i}.\label{R}
\end{equation}
Now the mentioned system of equations becomes:

\begin{equation}
\begin{cases}
\bm{\mathcal{I}}_{i=0}= & \mathbf{R}_{0}\left[\left(\mathbf{S}_{1}^{\mathrm{red}}-\mathbf{S}\right)\bm{\mathcal{I}}_{i=0}+\mathcal{I}_{1}^{\mathrm{el}}\mathbf{v}_{1}\right]\\
 & +\mathbf{R}_{-i_{2}}\left[\left(\mathbf{S}_{2}^{\mathrm{red}}-\mathbf{S}\right)\bm{\mathcal{I}}_{i_{2}}+\mathcal{I}_{2}^{\mathrm{el}}\mathbf{v}_{2}\right]\\
\bm{\mathcal{I}}_{i_{2}}= & \mathbf{R}_{i_{2}}\left[\left(\mathbf{S}_{1}^{\mathrm{red}}-\mathbf{S}\right)\bm{\mathcal{I}}_{i=0}+\mathcal{I}_{1}^{\mathrm{el}}\mathbf{v}_{1}\right]\\
 & +\mathbf{R}_{0}\left[\left(\mathbf{S}_{2}^{\mathrm{red}}-\mathbf{S}\right)\bm{\mathcal{I}}_{i_{2}}+\mathcal{I}_{2}^{\mathrm{el}}\mathbf{v}_{2}\right]
\end{cases}.\label{system of eq}
\end{equation}
Using block matrix notation system of equations \eqref{system of eq}
can be written as follows:

\begin{equation}
\mathbf{U}_{i_{2}}\begin{pmatrix}\bm{\mathcal{I}}_{i=0}\\
\bm{\mathcal{I}}_{i_{2}}
\end{pmatrix}=\mathcal{I}_{1}^{\mathrm{el}}\begin{pmatrix}\mathbf{R}_{0}\mathbf{v}_{1}\\
\mathbf{R}_{i_{2}}\mathbf{v}_{1}
\end{pmatrix}+\mathcal{I}_{2}^{\mathrm{el}}\begin{pmatrix}\mathbf{R}_{-i_{2}}\mathbf{v}_{2}\\
\mathbf{R}_{0}\mathbf{v}_{2}
\end{pmatrix},\label{eq:system of eq 2}
\end{equation}
where

\begin{equation}
\mathbf{U}_{i_{2}}=\begin{pmatrix}\mathbf{1}-\mathbf{R}_{0}\left(\mathbf{S}_{1}^{\mathrm{red}}-\mathbf{S}\right) & -\mathbf{R}_{-i_{2}}\left(\mathbf{S}_{2}^{\mathrm{red}}-\mathbf{S}\right)\\
-\mathbf{R}_{i_{2}}\left(\mathbf{S}_{1}^{\mathrm{red}}-\mathbf{S}\right) & \mathbf{1}-\mathbf{R}_{0}\left(\mathbf{S}_{2}^{\mathrm{red}}-\mathbf{S}\right)
\end{pmatrix}.\label{U matrix}
\end{equation}

Solving \eqref{eq:system of eq 2} leads to:

\begin{equation}
\begin{pmatrix}\bm{\mathcal{I}}_{i=0}\\
\bm{\mathcal{I}}_{i_{2}}
\end{pmatrix}=\mathcal{I}_{1}^{\mathrm{el}}\mathbf{U}_{i_{2}}^{-1}\begin{pmatrix}\mathbf{R}_{0}\mathbf{v}_{1}\\
\mathbf{R}_{i_{2}}\mathbf{v}_{1}
\end{pmatrix}+\mathcal{I}_{2}^{\mathrm{el}}\mathbf{U}_{i_{2}}^{-1}\begin{pmatrix}\mathbf{R}_{-i_{2}}\mathbf{v}_{2}\\
\mathbf{R}_{0}\mathbf{v}_{2}
\end{pmatrix}.
\end{equation}

Finally, we invoke the first two equations in Eq. \eqref{O=00003DSI extended}
and Eq. \eqref{S1}, to express $\mathcal{O}_{1}^{\mathrm{el}}$ and
$\mathcal{O}_{2}^{\mathrm{el}}$ in terms of $\mathcal{I}_{1}^{\mathrm{el}}$
and $\mathcal{I}_{2}^{\mathrm{el}}$. This gives:

\begin{align}
\mathbf{S}_{\mathrm{net}} & =\begin{pmatrix}\mathbf{w}_{1} & 0\\
0 & \mathbf{w}_{2}
\end{pmatrix}\mathbf{U}_{i_{2}}^{-1}\begin{pmatrix}\mathbf{R}_{0}\mathbf{v}_{1} & \mathbf{R}_{-i_{2}}\mathbf{v}_{2}\\
\mathbf{R}_{i_{2}}\mathbf{v}_{1} & \mathbf{R}_{0}\mathbf{v}_{2}
\end{pmatrix}\nonumber \\
 & +\begin{pmatrix}\left(\mathbf{S}_{1}\right)_{55} & 0\\
0 & \left(\mathbf{S}_{2}\right)_{55}
\end{pmatrix}.\label{Snet result!}
\end{align}

Formula \eqref{Snet result!} allows calculations of the net scattering
matrix, but gives rather a faint insight about its qualitative behavior.
This issue is addressed in the next section.

Inverse of the $\mathbf{U}$ matrix (which is $8\times8$) can be
expressed in terms of matrix expressions involving its $4\times4$
block matrices \citep{key-13} (by means of the Schur complement).
Although maximal matrix size is decreased, the final formulas are
by no means more concise and do not give any additional insight into
the problem.

In order to determine $\bm{\mathcal{I}}_{i}$, we do the inverse Fourier
transform on Eq. \eqref{Ik}. Substituting for $\bm{\mathcal{I}}_{i=0}$
and $\bm{\mathcal{I}}_{i_{2}}$ produces the following recipe:

\begin{align}
 & \bm{\mathcal{I}}_{i}=\nonumber \\
 & \left[\left.\left(\mathbf{R}_{i}\left(\mathbf{S}_{1}^{\mathrm{red}}-\mathbf{S}\right)\right|\mathbf{R}_{i-i_{2}}\left(\mathbf{S}_{2}^{\mathrm{red}}-\mathbf{S}\right)\right)\mathbf{U}_{i_{2}}^{-1}\right.\nonumber \\
 & \left.\times\begin{pmatrix}\mathbf{R}_{0}\mathbf{v}_{1} & \mathbf{R}_{-i_{2}}\mathbf{v}_{2}\\
\mathbf{R}_{i_{2}}\mathbf{v}_{1} & \mathbf{R}_{0}\mathbf{v}_{2}
\end{pmatrix}+\left.\left(\mathbf{R}_{i}\mathbf{v}_{1}\right|\mathbf{R}_{i-i_{2}}\mathbf{v}_{2}\right)\right]\begin{pmatrix}\mathcal{I}_{1}^{\mathrm{el}}\\
\mathcal{I}_{2}^{\mathrm{el}}
\end{pmatrix},\label{Ii}
\end{align}
where $\left.\left(\cdot\right|\cdot\right)$ denotes a row block
matrix. $\bm{\mathcal{O}}_{i}$ can be found directly from Eq. \eqref{Ii}
and Eq. \eqref{I in terms of O}.

The obtained formulas extensively use matrix inverses. Therefore,
it has been tacitly assumed that these inverses exist. Singularity
of $\mathbf{U}_{i_{2}}$ would reflect non-uniqueness of $\bm{\mathcal{I}}_{i=0}$
and $\bm{\mathcal{I}}_{i_{2}}$ possible due to bound states present
in the lattice. This problem does not affect $\mathbf{S}_{\mathrm{net}}$.
Taking the Moore-Penrose inverse of $\mathbf{U}_{i_{2}}$ removes
the issue. Singularity of $\mathbf{M}_{k}-\mathbf{S}$ for some $k$
is more problematic, since this matrix appears in Eq. \eqref{R} and
thus enters $\mathbf{U}_{i_{2}}$. It can be shown that singularity
of $\mathbf{M}_{k}-\mathbf{S}$ is equivalent to $\kappa$ corresponding
to an energy level from Sec. \ref{sec:Band-structure}. Namely, $\det\left(\mathbf{M}_{k}-\mathbf{S}\right)=0$
is equivalent to

\begin{equation}
0=\det\left[\left(\mathbf{M}_{k}-\mathbf{S}\right)\mathbf{M}_{k}\right]=\det\left[\mathbf{1}-\mathbf{S}\mathbf{M}_{k}\right],\label{0=00003Ddet}
\end{equation}
where unitarity of $\mathbf{M}_{k}$ has been used. Equation \eqref{0=00003Ddet}
can be rewritten as $\det\left[\mathbf{S}\mathbf{M}_{k}-\lambda\mathbf{1}\right]=0$,
with $\lambda=1$, which is exactly the condition found to indicate
whether $\kappa$ corresponds to an existing energy level in a pure
lattice (i. e. without external electrodes).

For a finite lattice, the set of allowed energies is discrete, so
the set of problematic (i. e. causing $\det\left(\mathbf{M}_{k}-\mathbf{S}\right)=0$
for some $k$) $\kappa$ values is of zero measure. This suggests
that the problem is rather apparent, although significance of singularity
of $\mathbf{M}_{k}-\mathbf{S}$ will reappear in the next section.

\begin{figure}
\centering{}\includegraphics[scale=0.65]{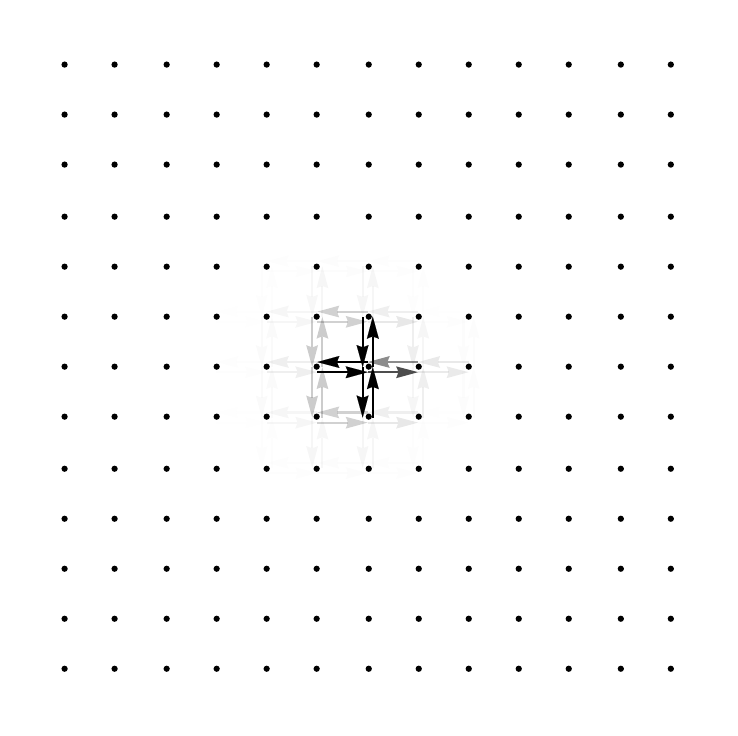}\vspace{-0.5cm}\caption{A mesoscopic analog of Fig. \ref{fig:Square-lattice}. Here, opacity
of arrows encodes squared amplitudes of the wave function propagating
in the waveguides.\label{fig:Mesoscopic-analog-of}}
\end{figure}

Figure \ref{fig:Mesoscopic-analog-of} shows graphically an output
of Eq. \eqref{Ii} for $\varphi_{1}=\pi/6$, $\varphi_{3}=-\pi/6$,
$\phi_{2}=0$, $d=1$, $\kappa=5.7$, $f=0$, $f_{2}=\kappa d$, $i_{2}=\left(1,0\right)$,
$\mathcal{I}_{1}^{\mathrm{el}}=1$ and $\mathcal{I}_{2}^{\mathrm{el}}=0$.

\section{Wave function patterns\label{sec:Wave-function-patterns}}

For the investigation of a wave function of a single injected electron,
the second electrode is not needed. It introduces additional impurity
to the model, which does not seem to be essential for the following
observations. Removing that electrode can be achieved by putting $\mathbf{v}_{2}=0,\mathbf{w}_{2}=0,\left(\mathbf{S}_{2}\right)_{55}=1$
and $\mathbf{S}_{2}^{\mathrm{red}}=\mathbf{S}$. This allows us to
focus on some interesting properties of the current spreading. Matrix
$\mathbf{U}_{i_{2}}$ reduces significantly and it is no longer needed.
Going back to Eq. \eqref{system of eq}, the system reduces to:

\begin{equation}
\begin{cases}
\left[\mathbf{1}-\mathbf{R}_{0}\left(\mathbf{S}_{1}^{\mathrm{red}}-\mathbf{S}\right)\right]\bm{\mathcal{I}}_{i=0} & =\mathcal{I}_{1}^{\mathrm{el}}\mathbf{R}_{0}\mathbf{v}_{1}\\
-\mathbf{R}_{i_{2}}\left(\mathbf{S}_{1}^{\mathrm{red}}-\mathbf{S}\right)\bm{\mathcal{I}}_{i=0}+\bm{\mathcal{I}}_{i_{2}} & =\mathcal{I}_{1}^{\mathrm{el}}\mathbf{R}_{i_{2}}\mathbf{v}_{1}
\end{cases}.\label{some eq}
\end{equation}

Eq. \eqref{some eq} is valid for any $i_{2}\neq0$, since after removing
the second electrode only site $i=0$ is special. Solving for $\bm{\mathcal{I}}_{i=0}$
from the first equation, we readily obtain $\bm{\mathcal{I}}_{i_{2}}$
(i. e. the complex amplitude distribution over all sites). Replacing
$i_{2}\rightarrow i$, the formula reads:

\begin{align}
 & \bm{\mathcal{I}}_{i}=\mathcal{I}_{1}^{\mathrm{el}}\mathbf{R}_{i}\nonumber \\
 & \times\left\{ \mathbf{1}+\left(\mathbf{S}_{1}^{\mathrm{red}}-\mathbf{S}\right)\left[\mathbf{1}-\mathbf{R}_{0}\left(\mathbf{S}_{1}^{\mathrm{red}}-\mathbf{S}\right)\right]^{-1}\mathbf{R}_{0}\right\} \mathbf{v}_{1}.\label{Ii2}
\end{align}

\begin{figure}
\centering{}\includegraphics[scale=0.4]{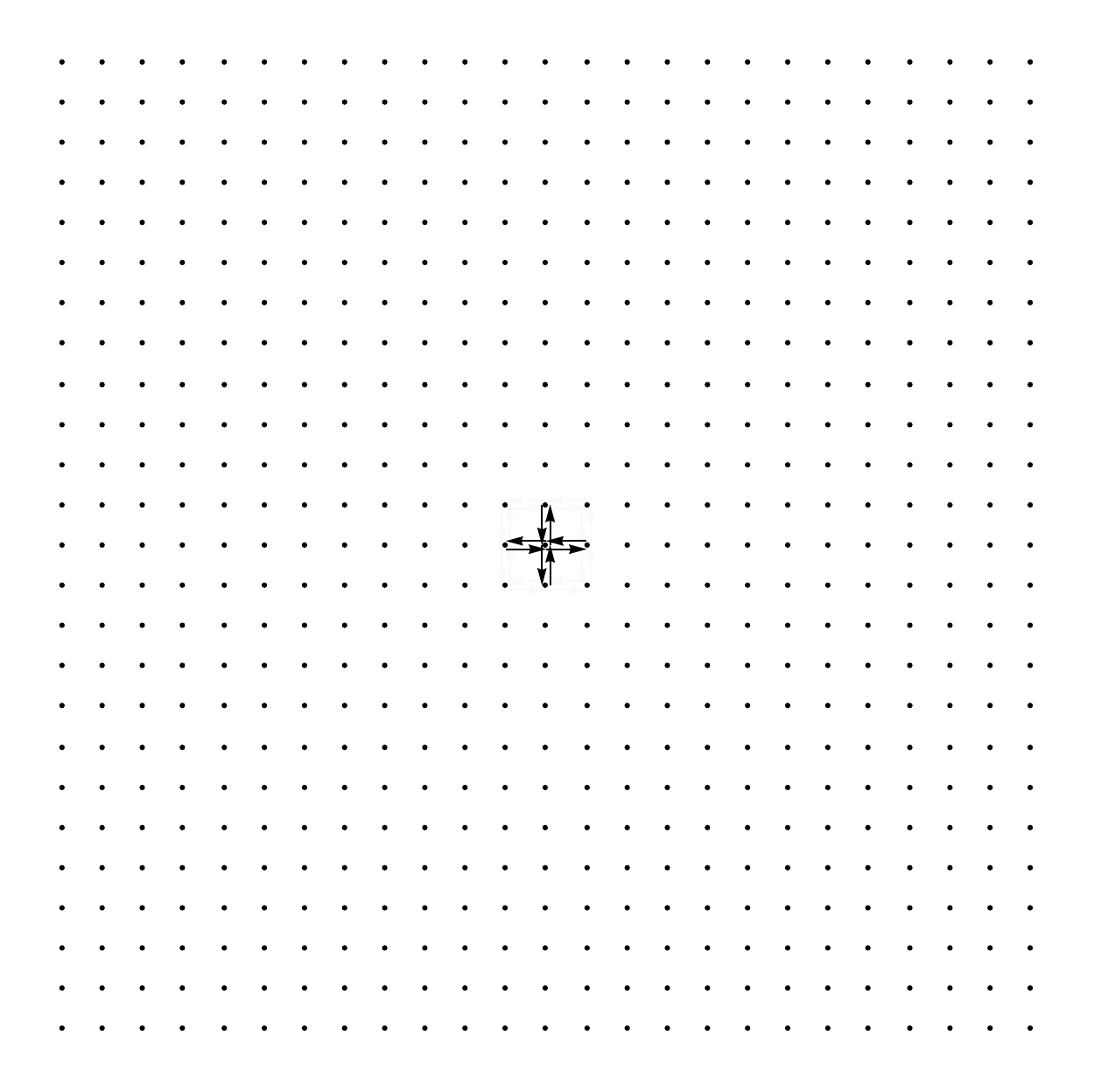}\caption{$\kappa=4.3$\label{fig: kappa=00003D4.3}}
\end{figure}

\begin{figure}
\centering{}\includegraphics[scale=0.4]{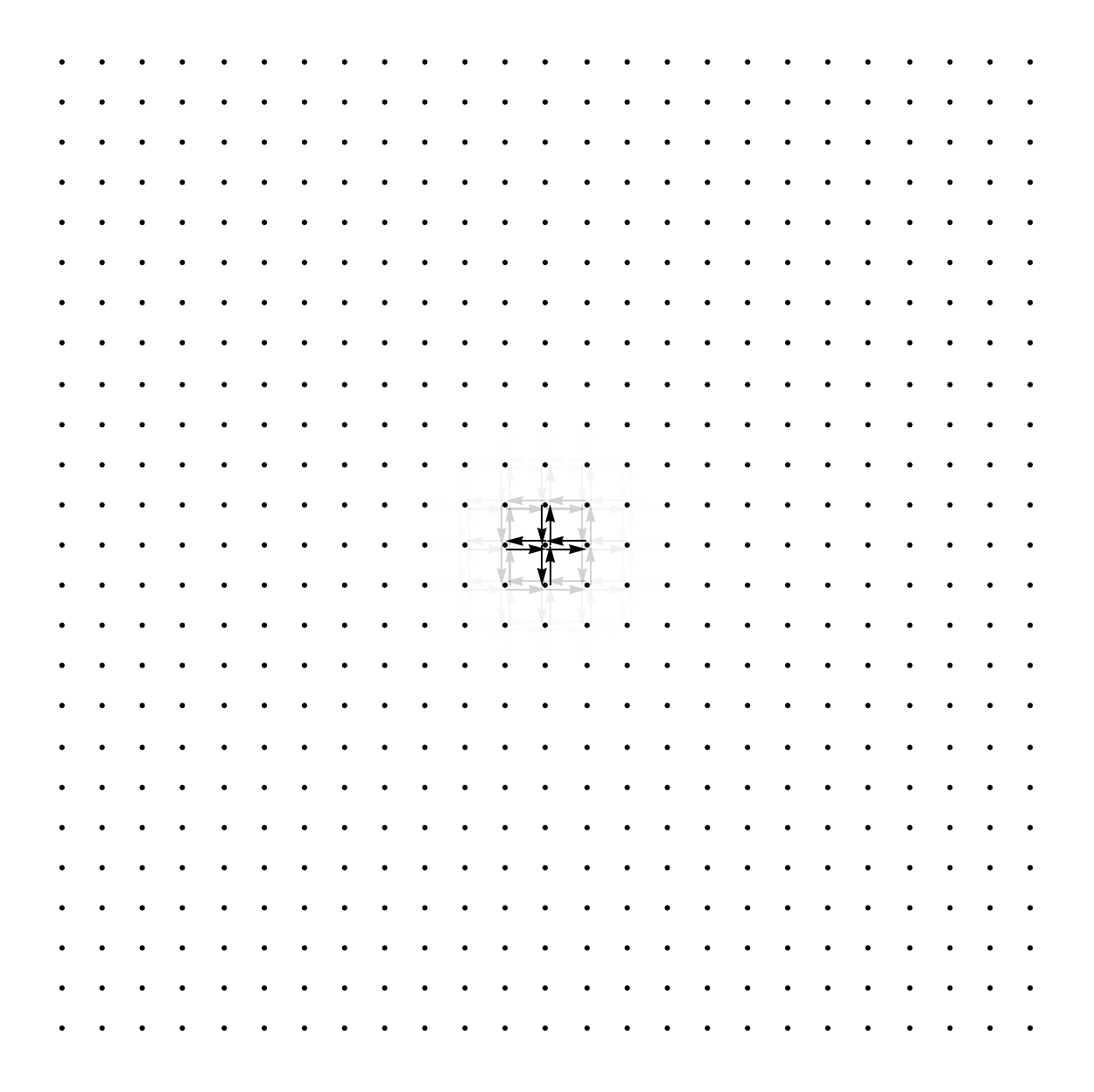}\caption{$\kappa=5.7$\label{fig: kappa=00003D5.7}}
\end{figure}

\begin{figure}
\centering{}\includegraphics[scale=0.4]{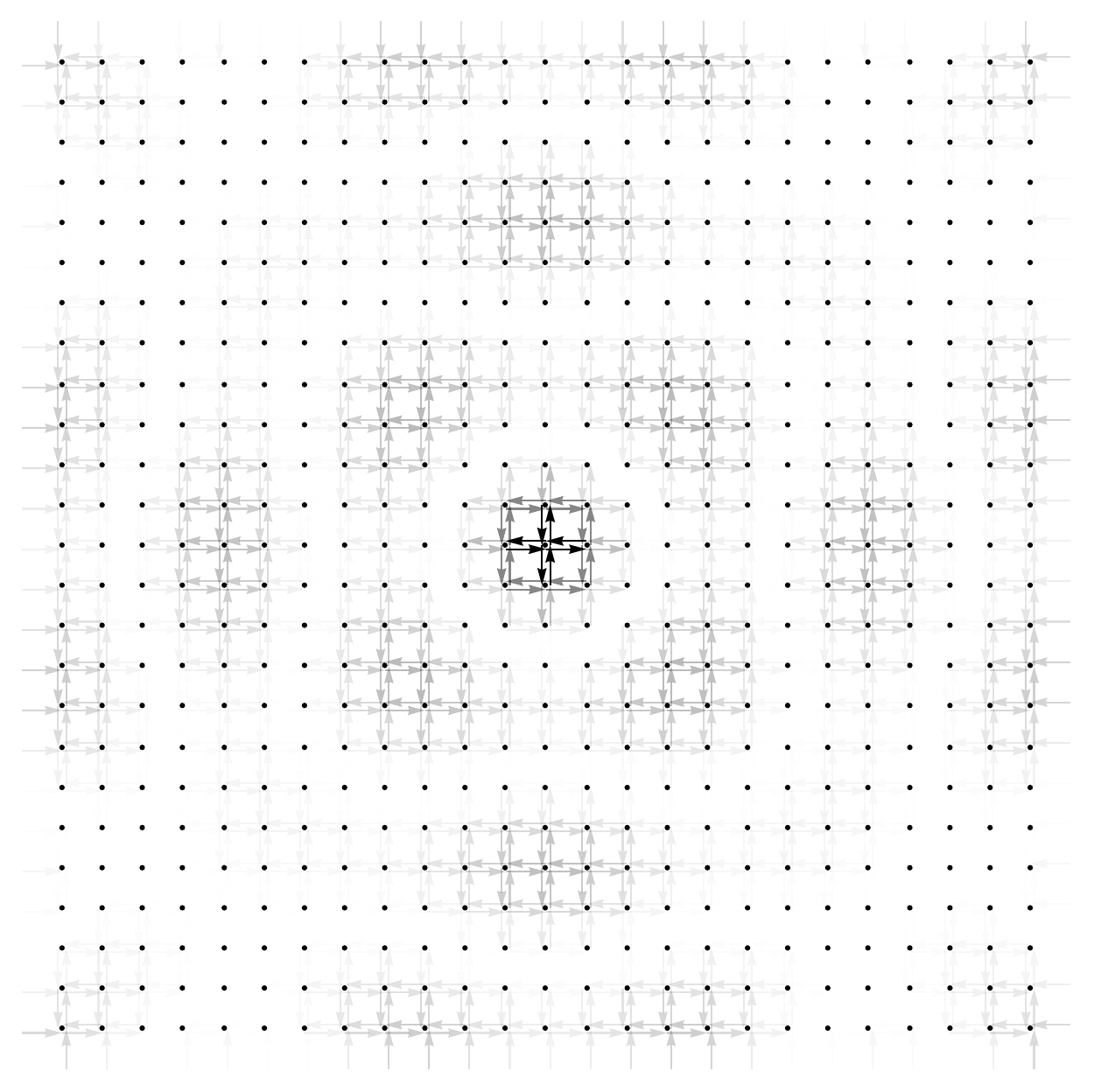}\caption{$\kappa=5.8$\label{fig: kappa=00003D5.8}}
\end{figure}

\begin{figure}
\centering{}\includegraphics[scale=0.4]{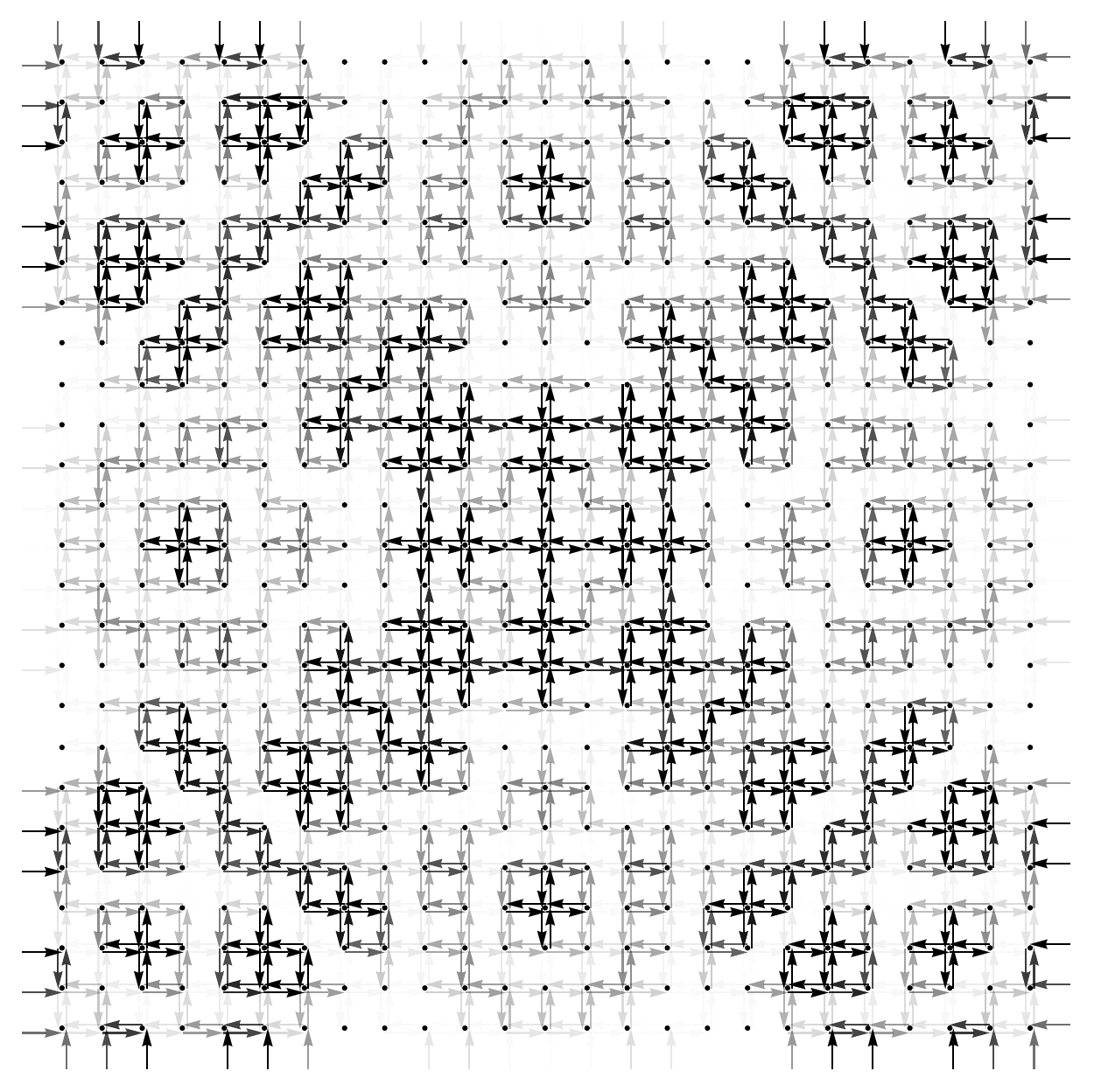}\caption{$\kappa=5.9$\label{fig: kappa=00003D5.9}}
\end{figure}

\begin{figure*}
\centering{}\subfloat[$\kappa=6.0$]{\centering{}\includegraphics[scale=0.43]{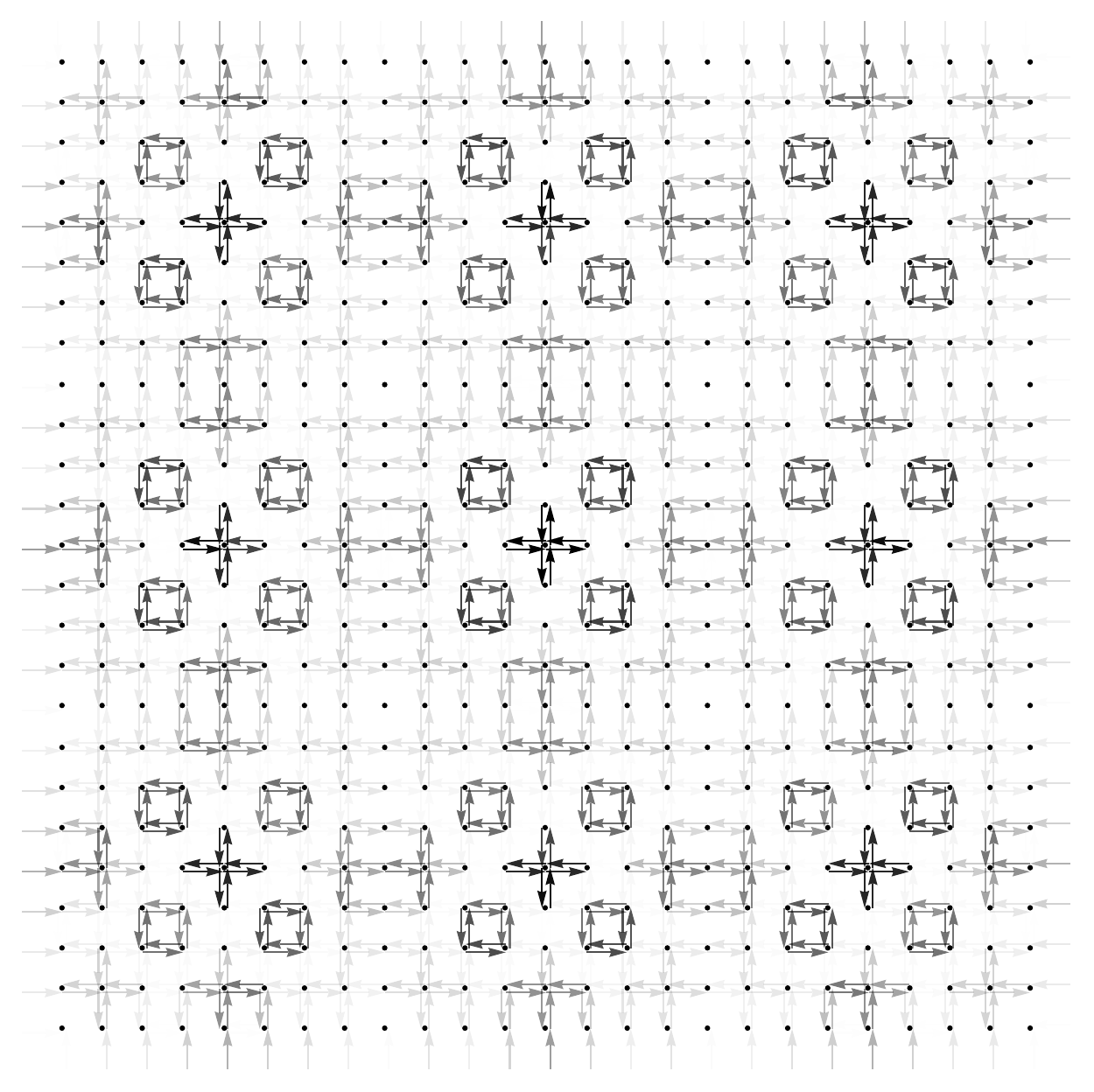}}\\\subfloat[$\kappa=6.2$]{\centering{}\includegraphics[scale=0.43]{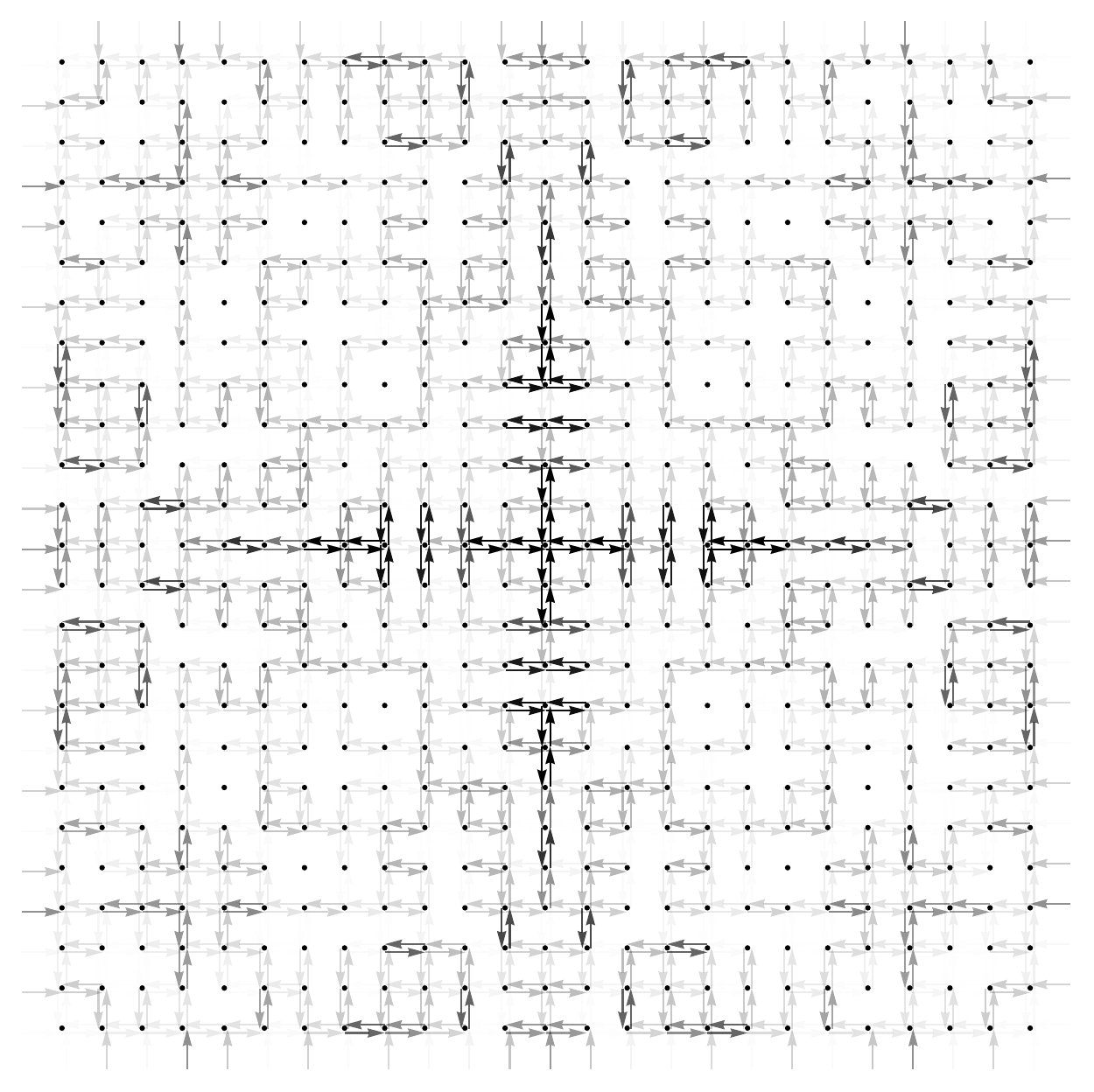}}\caption{Two another intricate examples\label{fig:Two-another-cool}}
\end{figure*}

With fixed $\varphi_{1}=\pi/6$, $\varphi_{3}=-\pi/6$, $\phi_{2}=0$,
$d=1$, $f=0$ and $f_{2}=\kappa d$, $\kappa$ was varied and Eq.
\eqref{Ii2} was used to create patterns in the style of Fig. \ref{fig:Mesoscopic-analog-of}.
For the first time, with $\kappa=4.3$, a rather dull structure was
observed (Fig. \ref{fig: kappa=00003D4.3}). Sadly, it seemed stable
upon changing $\kappa$. At some point ($\kappa=5.7$, Fig. \ref{fig: kappa=00003D5.7}),
the current distribution started to grow slightly. Subsequently ($\kappa=5.8$,
Fig. \ref{fig: kappa=00003D5.8}), significant current appeared far
from the electrode – the electron was no longer confined in the small
central area. At $\kappa=5.9$ (Fig. \ref{fig: kappa=00003D5.9})
the electron is completely delocalized. Many different intriguing
patterns were observed. Two different examples are given in Fig. \ref{fig:Two-another-cool}.
Intricate structures persisted until roughly $\kappa=6.8$, when again
clear localization around the electrode appeared.

A natural question arises: Can we predict whether a localized or intricate
pattern occurs (without determining it in detail)? A formal way of
recognizing localized patterns can be by examining the asymptotic
behavior of $\bm{\mathcal{I}}_{i}$ as $i$ tends to infinity. The
only site-dependent term in Eq. \eqref{Ii2} is $\mathbf{R}_{i}$,
so we can focus on asymptotic behavior of its elements.

It is a moment when some clarification regarding the size of the lattice
is needed. All condensed-matter physical systems are finite and specifically
the grid of mesoscopic resistors has to be as well (moreover, its
size must be still much smaller than the coherence length). Therefore,
the word ``infinite'' in the title of this paper means ``large'',
or more precisely, means that we are interested in some properties
formally present in the limit of infinite size. For example, band
structure from Fig. \ref{fig:Band-structure} and \ref{fig:The-lowest-three}
becomes continuous only in this limit (which from now will be referred
to as thermodynamic). All wave function visualizations were performed
for a finite, but large (namely $64\times64$) lattice. A standard
thermodynamic limit substitution turning a sum into an integral ($\frac{1}{N}\sum_{k}\rightarrow\frac{1}{\left(2\pi\right)^{2}}\int\mathrm{d}^{2}k$)
leads to the following expression for $\mathbf{R}_{i}$:

\begin{equation}
\mathbf{R}_{i}=\frac{1}{\left(2\pi\right)^{2}}\int_{-\pi}^{\pi}\int_{-\pi}^{\pi}\mathrm{d}k_{x}\mathrm{d}k_{y}\left(\mathbf{M}_{k}-\mathbf{S}\right)^{-1}e^{\mathrm{i}k\cdot i}.\label{Ri Int}
\end{equation}
However, this formula breaks when $\left(\mathbf{M}_{k}-\mathbf{S}\right)^{-1}$
has singularities. Let us first focus on a simpler case, when such
singularities do not occur for any $k\in\left(-\pi,\pi\right]\times\left(-\pi,\pi\right]$.

Elements of $\left(\mathbf{M}_{k}-\mathbf{S}\right)^{-1}$ are rational
functions of $e^{\mathrm{i}k_{x}}$, $e^{-\mathrm{i}k_{x}}$, $e^{\mathrm{i}k_{y}}$,
$e^{-\mathrm{i}k_{y}}$. Focusing on just one direction (e. g. $x$),
any element of $\mathbf{R}_{i}$ can be written as

\begin{equation}
\frac{1}{\left(2\pi\right)^{2}}\int_{-\pi}^{\pi}\mathrm{d}k_{y}\,I\left(k_{y}\right)e^{\mathrm{i}k_{y}i_{y}},
\end{equation}
where

\begin{equation}
I\left(k_{y}\right)=\int_{-\pi}^{\pi}\mathrm{d}k_{x}\frac{P_{k_{y}}\left(e^{\mathrm{i}k_{x}},e^{-\mathrm{i}k_{x}}\right)}{Q_{k_{y}}\left(e^{\mathrm{i}k_{x}},e^{-\mathrm{i}k_{x}}\right)}e^{\mathrm{i}k_{x}i_{x}},
\end{equation}
with $P_{k_{y}}$ and $Q_{k_{y}}$ representing some polynomials of
$e^{\mathrm{i}k_{x}},e^{-\mathrm{i}k_{x}}$ with coefficients dependent
on $k_{y}$. Introducing a complex variable $z=e^{\mathrm{i}k_{x}}$,
a $\int_{-\pi}^{\pi}\mathrm{d}k_{x}$ integral can be exchanged for
a complex integral over a unit circle:

\begin{equation}
I\left(k_{y}\right)=\oint\frac{\mathrm{d}z}{\mathrm{i}z}\frac{P_{k_{y}}\left(z,z^{-1}\right)}{Q_{k_{y}}\left(z,z^{-1}\right)}z^{i_{x}}.
\end{equation}

Since $P$ and $Q$ are finite bivariate polynomials, there exist
such integers $p$ and $q$, that $z^{p}P_{k_{y}}\left(z,z^{-1}\right)=\tilde{P}_{k_{y}}\left(z\right)$
and $z^{q}Q_{k_{y}}\left(z,z^{-1}\right)=\tilde{Q}_{k_{y}}\left(z\right)$
are polynomials in $z$ (with no $z=0$ roots and also, which can
be assumed without loss of generality, no common roots). Thus:

\begin{equation}
I\left(k_{y}\right)=-\mathrm{i}\oint\mathrm{d}z\,\frac{\tilde{P}_{k_{y}}\left(z\right)}{\tilde{Q}_{k_{y}}\left(z\right)}z^{i_{x}+q-p-1}.
\end{equation}

Polynomial $\tilde{Q}_{k_{y}}\left(z\right)$ can be factorized as
$\prod_{\ell}\left(z-z_{\ell}\right)^{m_{\ell}}$ (unless it is constant,
but then for sufficiently large $i_{x}$ integral $I$ would be zero).
$\ell$ indexes roots of $\tilde{Q}$ and $m_{\ell}$ denotes their
multiplicity. By assumption, $\left(\mathbf{M}_{k}-\mathbf{S}\right)^{-1}$
has no singularities, so $\left|z_{\ell}\right|\neq1$ for all $\ell$.
Let $A$ be the set of all such indices $\ell$ that $\left|z_{\ell}\right|<1$.
Partitioning the unit circle into many counterclockwise contours (each
encircling only one root), $I$ takes the form

\begin{align}
I\left(k_{y}\right) & =-\mathrm{i}\sum_{\ell\in A}\underset{\text{only }z_{\ell}}{\underset{\text{contour enclosing}}{\underset{\text{counterclockwise}}{\oint}}}\mathrm{d}z\nonumber \\
 & \times\frac{\tilde{P}_{k_{y}}\left(z\right)/\prod_{\ell^{\prime}\neq\ell}\left(z-z_{\ell^{\prime}}\right)^{m_{\ell^{\prime}}}}{\left(z-z_{\ell}\right)^{m_{\ell}}}z^{i_{x}+q-p-1}.
\end{align}

Using the Cauchy integral formula, we get:

\begin{align}
 & I\left(k_{y}\right)=\sum_{\ell\in A}\frac{2\pi}{\left(m_{\ell}-1\right)!}\nonumber \\
 & \times\left[\frac{\mathrm{d}^{m_{\ell}-1}}{\mathrm{d}z^{m_{\ell}-1}}\left(\frac{\tilde{P}_{k_{y}}\left(z\right)}{\prod_{\ell^{\prime}\neq\ell}\left(z-z_{\ell^{\prime}}\right)^{m_{\ell^{\prime}}}}z^{i_{x}+q-p-1}\right)\right]_{z=z_{\ell}}.
\end{align}

We are interested in the behavior of $I$ for $i_{x}\gg1$. Therefore,
focusing only on asymptotic form:

\begin{equation}
I\left(k_{y}\right)\cong\sum_{\ell\in A}c_{\ell}i_{x}^{m_{\ell}-1}z_{\ell}^{i_{x}},\label{exp decay}
\end{equation}
where $c_{\ell}$ are some constants. The main conclusion is, that
since $\left|z_{\ell}\right|<1$ for $\ell\in A$, $I\left(k_{y}\right)$
(and thus $\mathbf{R}_{i}$) decays exponentially with $i_{x}$.

It is clear, how the given argument gets spoiled by singularities
acquired by $\left(\mathbf{M}_{k}-\mathbf{S}\right)^{-1}$. They introduce
roots $z_{\ell}$ lying on the unit circle, which ruins complex integral
evaluation and also circumvents final observation concerning Eq. \eqref{exp decay}.
In such case, it is not even obvious whether $\mathbf{R}_{i}$ approaches
anything in the thermodynamic limit. Fourier transforms of functions
with poles may be non-vanishing for $\left|i\right|\rightarrow\infty$,
as exemplified by the principal value of a one dimensional integral
$\int_{-\infty}^{\infty}\mathrm{d}k\frac{e^{\mathrm{i}ki}}{k}=\mathrm{i}\pi$.

Therefore, intricate patterns (for sufficiently large lattices) can
only occur (and generally do occur) when $\left(\mathbf{M}_{k}-\mathbf{S}\right)^{-1}$
(as a continuous function of $k\in\left(-\pi,\pi\right]\times\left(-\pi,\pi\right]$)
has singularities. As justified by Eq. \eqref{0=00003Ddet}, this
condition is equivalent to matching a band with the energy of the
injected electron. It is possible to derive a closed-form condition
for the presence of an intricate pattern, which involves directly
the relevant parameters. Details are given in \ref{sec:Derivation-of-the-1}.

\section{Experimental realization and conclusion\label{sec:Experimental-realization-and}}

Although the entire paper was propelled by the idea of bringing a
classical (and classic) infinite grid of resistors to a mesoscopic
scale, it is really scattering and interference, that have been truly
investigated. Besides this former realization, which is rather hard
to achieve, many other systems can be described by the considered
model. Instead of an electron wave function, light or acoustic waves
could be used.

In the first case, optical fibers can act as quantum wires and four-port
star couplers as junctions. The electrodes would be implemented as
additional fibers joined to five-port star couplers. Feeding the system
with laser light would generate the considered interference pattern
inside it. Observing it would require some additional fibers making
the light leak out of the grid (i. e. weakly coupled to the system)
and shine on a screen. A different alternative might be replacing
fibers by free propagation. Then presence of some dust in the air
would weakly scatter the light making the intensity distribution directly
visible. However, this solution rises technical difficulties with
scattering the light by means of star couplers.

In the acoustic case, the mesoscopic grid of resistors gets replaced
by a grid of pipes. Visualization of the interference pattern can
be achieved by means of the solution from the Rubens tube \citep{key-14}.
Flames besides indicating nodes and antinodes would also characterize
the amplitude in the latter.

It should be noted that contributions from different modes (not analyzed
in the model) may affect the experiment slightly, as it is the case
in multi-mode fiber optics or the classic Rubens tube demonstration.

Waves proved to be promising ornament designers, so it may be an intriguing
artistic experience to watch in real time the evolution of the interference
pattern. In case of laser light, it would be accompanied by color
changes, while in the acoustic case – by pitch changes.

Achieving a transition from an intricate to a localized pattern driven
by wavelength changes requires a gapped spectrum. This condition,
as seen from Eq. \eqref{extrema}, depends on phases $\phi_{1},\phi_{2},\phi_{3}$
introduced in Eq. \eqref{bfs}. As shown in \citep{key-8}, a gapless
spectrum is possible for certain values of these phases, so it is
of interest how to avoid such a situation. Phases $\phi_{1}$, $\phi_{2}$,
$\phi_{3}$ and $f_{1}$, $f_{2}$ depend on the detailed geometry
of the junctions. It is an interesting question whether any desired
values of these can be achieved experimentally, but the answer is
not obvious to the author. However, it is sufficient to know if phases
generating a gapped spectrum are easily obtainable. The following
heuristic argument suggests that it is the case. Considering an almost
fully backscattering junctions, parameters from equation (9) are $b=-1$,
$s=f=0$. This gives phases $\phi_{1}=\phi_{2}=\phi_{3}=\pi$. Then
Eq. \eqref{extrema} predicts a localized pattern for any $\kappa$
(excluding a set of zero measure). If a junction unluckily leads to
a gapless spectrum, redesigning it for more backscattering should
fix the issue. Of course, determining the phases precisely for a given
junction would require a separate numerical analysis.

Beyond purely esthetic and educational value, the main finding of
the paper, that delocalized intricate interference patterns correspond
to fitting the wavelength into a band, could be used to investigate
experimentally band structures of similar systems.

\appendix

\section{Derivation of the analytical dispersion relation\label{sec:Derivation-of-the}}

Assuming $\varphi_{3}=-\varphi_{1}$, the characteristic polynomial
of matrix $\mathbf{m}_{k}$ equated to zero reads:

\begin{align}
 & \lambda^{4}+\left(1-\cos\varphi_{1}\right)\left(\cos k_{x}+\cos k_{y}\right)\lambda^{3}\nonumber \\
 & +2\left[\cos k_{x}\cos k_{y}\left(1-\cos\varphi_{1}\right)-\cos\varphi_{1}\right]\lambda^{2}\nonumber \\
 & +\left(1-\cos\varphi_{1}\right)\left(\cos k_{x}+\cos k_{y}\right)\lambda+1=0.\label{poly}
\end{align}

All the coefficients are real, so the roots will appear in pairs $\lambda,\lambda^{*}$.
Additionally, unitarity of $\mathbf{m}_{k}$ enforces all the roots
to lie on the unit circle in the complex plane. Thus we can write
them as $e^{\mathrm{i}u_{1}},e^{-\mathrm{i}u_{1}},e^{\mathrm{i}u_{2}},e^{-\mathrm{i}u_{2}}$.
However, if $\lambda=\pm1$ was a single root, then its complex conjugate
would be identical to it, and the mentioned set of roots would be
incomplete. In such case, both $\lambda=1$ and $\lambda=-1$ would
have to be roots, because otherwise their total number would be odd.
Checking the values of the polynomial at $\lambda=1$ and $\lambda=-1$
gives $16\cos^{2}\frac{k_{x}}{2}\cos^{2}\frac{k_{y}}{2}\sin^{2}\frac{\varphi_{1}}{2}$
and $2\left(1-\cos k_{x}\right)\left(1-\cos k_{y}\right)\left(1-\cos\varphi_{1}\right)$.
Assuming $\varphi_{1}\neq0$ (and $\varphi_{1}\in\left(-\pi,\pi\right]$),
they can be simultaneously $0$ only on a zero-measure subset of the
$k$-space. Thus we ignore this case and continue assuming all the
roots can be written as $e^{\mathrm{i}u_{1}},e^{-\mathrm{i}u_{1}},e^{\mathrm{i}u_{2}},e^{-\mathrm{i}u_{2}}$.

Now we can compare the factored form of the characteristic polynomial

\begin{equation}
\left(\lambda-e^{\mathrm{i}u_{1}}\right)\left(\lambda-e^{-\mathrm{i}u_{1}}\right)\left(\lambda-e^{\mathrm{i}u_{2}}\right)\left(\lambda-e^{-\mathrm{i}u_{2}}\right),
\end{equation}
with its expanded form (the left-hand-side of Eq. \eqref{poly}).
This leads to the following system of equations:

\begin{equation}
\begin{cases}
\cos u_{1}+\cos u_{2}=-\frac{\left(1-\cos\varphi_{1}\right)\left(\cos k_{x}+\cos k_{y}\right)}{2}\\
\cos u_{1}\cos u_{2}=\frac{\cos k_{x}\cos k_{y}\left(1-\cos\varphi_{1}\right)-\cos\varphi_{1}-1}{2}
\end{cases}.
\end{equation}

Writing it as

\begin{equation}
\begin{cases}
x_{1}+x_{2} & =\sigma\\
x_{1}x_{2} & =p
\end{cases},
\end{equation}
reveals that its structure is identical to Vieta's formulas for the
quadratic equation. Thus $x_{1}=\frac{\sigma}{2}-\sqrt{\frac{\sigma^{2}}{4}-p},\:x_{2}=\frac{\sigma}{2}+\sqrt{\frac{\sigma^{2}}{4}-p}$.
Finally:

\begin{equation}
u_{1}=\arccos\left(\frac{\sigma}{2}-\sqrt{\frac{\sigma^{2}}{4}-p}\right),
\end{equation}

\begin{equation}
u_{2}=\arccos\left(\frac{\sigma}{2}+\sqrt{\frac{\sigma^{2}}{4}-p}\right),
\end{equation}
which leads directly to Eq. \eqref{E}.

\section{Derivation of the analytical intricacy condition\label{sec:Derivation-of-the-1}}

$\left(\mathbf{M}_{k}-\mathbf{S}\right)^{-1}$ has a singularity,
when $\det\left(\mathbf{M}_{k}-\mathbf{S}\right)=0$. Symbolic evaluation
of the determinant leads to:

\begin{align}
 & \det\left(\mathbf{M}_{k}-\mathbf{S}\right)=-e^{\mathrm{i}\left(2d\kappa+\frac{\varphi_{1}}{2}+\frac{\varphi_{3}}{2}+2\text{\ensuremath{\phi_{2}}}\right)}\nonumber \\
 & \times\left\{ -2\cos\left(2d\kappa+\frac{\varphi_{1}}{2}+\frac{\varphi_{3}}{2}+2\phi_{2}\right)\right.\nonumber \\
 & +\left(\cos k_{x}+\cos k_{y}\right)\left[\cos\left(d\kappa+\frac{\varphi_{1}}{2}-\frac{\varphi_{3}}{2}+\phi_{2}\right)\right.\nonumber \\
 & +\cos\left(d\kappa-\frac{\varphi_{1}}{2}+\frac{\varphi_{3}}{2}+\phi_{2}\right)\nonumber \\
 & \left.-2\cos\left(d\kappa+\frac{\varphi_{1}}{2}+\frac{\varphi_{3}}{2}+\phi_{2}\right)\right]\nonumber \\
 & +4\cos k_{x}\cos k_{y}\sin\left(\frac{\text{\ensuremath{\varphi_{1}}}}{2}\right)\sin\left(\frac{\text{\ensuremath{\varphi_{3}}}}{2}\right)\nonumber \\
 & \left.+2\cos\left(\frac{\varphi_{1}-\varphi_{3}}{2}\right)\right\} .\label{det MKS}
\end{align}

$\det\left(\mathbf{M}_{k}-\mathbf{S}\right)$ times $e^{-\mathrm{i}\left(2d\kappa+\frac{\varphi_{1}}{2}+\frac{\varphi_{3}}{2}+2\text{\ensuremath{\phi_{2}}}\right)}$
is a real function of $k$, which has the following form:

\begin{equation}
f\left(k_{x},k_{y}\right)=A\cos k_{x}\cos k_{y}+B\left(\cos k_{x}+\cos k_{y}\right)+C.\label{form}
\end{equation}

We show that a global extremum occurs only for $\sin k_{x}=\sin k_{y}=0$
(assuming $A\neq B$; for $A=B$ one global extremum is at $\left(0,0\right)$
and the other extends on the boundary of the $\left[-\pi,\pi\right]\times\left[-\pi,\pi\right]$
square). Partial derivatives of $f$ equated to zero read:

\begin{equation}
\begin{cases}
\sin k_{x}\left(A\cos k_{y}+B\right)=0\\
\sin k_{y}\left(A\cos k_{x}+B\right)=0
\end{cases}.\label{eq sys}
\end{equation}

The Hessian matrix of $f$ is given by:

\begin{align}
 & H=\nonumber \\
 & \begin{pmatrix}-\cos k_{x}\left(A\cos k_{y}+B\right) & A\sin k_{x}\sin k_{y}\\
A\sin k_{x}\sin k_{y} & -\cos k_{y}\left(A\cos k_{x}+B\right)
\end{pmatrix}.
\end{align}

Now all candidates for extrema are examined.

If $A\cos k_{y}+B=0$ or $A\cos k_{x}+B=0$, then $\det H=-A^{2}\sin^{2}k_{x}\sin^{2}k_{y}$.
If both $\sin k_{x}$ and $\sin k_{y}$ are nonzero, then $\det H<0$
and this corresponds to a saddle-point. If exactly one of $\sin k_{x}$
and $\sin k_{y}$ is zero, let us say $\sin k_{x}$, then $f$ becomes
a linear function of $\cos k_{y}$ having an extremum at $k_{y}=0$
or $k_{y}=\pm\pi$ (so $\sin k_{x}=\sin k_{y}=0$).

If $A\cos k_{y}+B\neq0$ and $A\cos k_{x}+B\neq0$, then Eq. \eqref{eq sys}
enforces that $\sin k_{x}=\sin k_{y}=0$.

For $\sin k_{x}=\sin k_{y}=0$ we have to consider only $k$ being
$0$ or $\pi$ (since $-\pi$ and $\pi$ are equivalent).

It turns out that:

\begin{align}
 & -\frac{e^{-\mathrm{i}\left(2d\kappa+\frac{\varphi_{1}}{2}+\frac{\varphi_{3}}{2}+2\text{\ensuremath{\phi_{2}}}\right)}}{16}\det\left(\mathbf{M}_{\left(0,0\right)}-\mathbf{S}\right)\nonumber \\
 & =\cos^{2}\left(\frac{d\kappa+\phi_{2}}{2}\right)\sin\left(\frac{d\kappa+\varphi_{1}+\phi_{2}}{2}\right)\nonumber \\
 & \times\sin\left(\frac{d\kappa+\varphi_{3}+\phi_{2}}{2}\right),
\end{align}

\begin{align}
 & -\frac{e^{-\mathrm{i}\left(2d\kappa+\frac{\varphi_{1}}{2}+\frac{\varphi_{3}}{2}+2\text{\ensuremath{\phi_{2}}}\right)}}{16}\det\left(\mathbf{M}_{\left(\pi,\pi\right)}-\mathbf{S}\right)\nonumber \\
 & =\sin^{2}\left(\frac{d\kappa+\phi_{2}}{2}\right)\cos\left(\frac{d\kappa+\varphi_{1}+\phi_{2}}{2}\right)\nonumber \\
 & \times\cos\left(\frac{d\kappa+\varphi_{3}+\phi_{2}}{2}\right),
\end{align}

\begin{align}
 & -\frac{e^{-\mathrm{i}\left(2d\kappa+\frac{\varphi_{1}}{2}+\frac{\varphi_{3}}{2}+2\text{\ensuremath{\phi_{2}}}\right)}}{16}\det\left(\mathbf{M}_{\left(0,\pi\right)}-\mathbf{S}\right)\nonumber \\
 & =\frac{1}{4}\sin\left(d\kappa+\phi_{2}\right)\sin\left(d\kappa+\frac{\varphi_{1}}{2}+\frac{\varphi_{3}}{2}+\phi_{2}\right).
\end{align}

According to Eq. \eqref{det MKS},
\begin{equation}
-\frac{1}{16}\det\left(\mathbf{M}_{k}-\mathbf{S}\right)e^{-\mathrm{i}\left(2d\kappa+\frac{\varphi_{1}}{2}+\frac{\varphi_{3}}{2}+2\text{\ensuremath{\phi_{2}}}\right)}\label{expr}
\end{equation}
is a real function of $k$, being of the form given by Eq. \eqref{form}.
All candidates for global extrema are:

\begin{equation}
\begin{cases}
\cos^{2}\left(\frac{d\kappa+\phi_{2}}{2}\right)\sin\left(\frac{d\kappa+\varphi_{1}+\phi_{2}}{2}\right)\sin\left(\frac{d\kappa+\varphi_{3}+\phi_{2}}{2}\right)\\
\sin^{2}\left(\frac{d\kappa+\phi_{2}}{2}\right)\cos\left(\frac{d\kappa+\varphi_{1}+\phi_{2}}{2}\right)\cos\left(\frac{d\kappa+\varphi_{3}+\phi_{2}}{2}\right)\\
\frac{1}{4}\sin\left(d\kappa+\phi_{2}\right)\sin\left(d\kappa+\frac{\varphi_{1}}{2}+\frac{\varphi_{3}}{2}+\phi_{2}\right)
\end{cases}.\label{extrema}
\end{equation}

All values taken by Eq. \eqref{expr} belong to the interval between
the smallest and the biggest number from Eq. \eqref{extrema}. Thus
the condition for $\det\left(\mathbf{M}_{k}-\mathbf{S}\right)$ to
be $0$ for some $k$, is that not all the values in \eqref{extrema}
be of the same sign (unless they are all zero). Therefore, this is
the condition (which is strictly speaking a necessary condition) for
the intricacy of the wave function.

\end{document}